\documentclass[pre,tighten]{revtex4-1}

\usepackage{amssymb,amsmath}

\newcommand\beq{\begin{equation}}
\newcommand\eeq{\end{equation}}
\newcommand\beqa{\begin{eqnarray}}
\newcommand\eeqa{\end{eqnarray}}

\usepackage{graphicx}

\usepackage{amsmath}
\usepackage{amssymb}
\usepackage{bm}
\usepackage{mathrsfs}
\usepackage{color}

\makeatletter

\def\btt#1{\texttt{\@backslashchar#1}}%

\DeclareRobustCommand\bblash{\btt{\@backslashchar}}%
\makeatother

\begin{document}

\title{Kinetic theory of shear thickening for a moderately dense gas-solid suspension:
from discontinuous thickening to continuous thickening}
\author{Hisao Hayakawa}\email[e-mail:]{hisao@yukawa.kyoto-u.ac.jp}
\affiliation{Yukawa Institute for Theoretical Physics, Kyoto University, Kyoto 606-8502, Japan}
\author{Satoshi Takada}
\affiliation{Earthquake Research Institute, The University of Tokyo, 1-1-1, Yayoi, Bunkyo-ku, Tokyo, 113-0032 Japan}
\affiliation{Department of Physics, Kyoto University, Kyoto 606-8502, Japan}
\author{Vicente Garz\'{o}}
\affiliation{
Departamento de F\'{\i}sica and Instituto de Computaci\'on Cient\'{\i}fica Avanzada (ICCAEX),
Universidad de Extremadura, E-06071 Badajoz, Spain
}

\begin{abstract}

The Enskog kinetic theory for moderately dense gas-solid suspensions under simple shear flow is considered as a model to analyze the rheological properties of the system. The influence of the environmental fluid on solid particles is modeled via a viscous drag force plus a stochastic Langevin-like term. 
The Enskog equation is solved by means of two independent but complementary routes: (i) Grad's moment method and (ii) event-driven Langevin simulation of hard spheres. Both approaches clearly show
that the flow curve (stress-strain rate relation) depends significantly on the volume fraction of the solid particles. In particular, as the density increases, there is a transition from the discontinuous shear thickening (observed in dilute gases) to the continuous shear thickening for denser systems. The comparison between theory and simulations indicate that while the theoretical predictions for the kinetic temperature agree well with simulations for densities $\varphi \lesssim 0.5$, the agreement for the other rheological quantities (the viscosity, the stress ratio and the normal stress differences) is limited to more moderate densities ($\varphi \lesssim 0.3$) if the inelasticity during collisions between particles is not large.

[This paper has been published in Phys. Rev. E {\bf 96}, 42903 (2017) but we have realized that there are some typos and mistakes after its publication. So we add the Erratum in the end of this paper.] 

\end{abstract}
\date{\today}

\maketitle

\section{Introduction}
\label{introduction}

Shear thickening is a rheological process in which the viscosity increases as the shear rate increases.
There are two types of shear thickenings, the continuous shear thickening (CST) and
 the discontinuous shear thickening (DST).
In particular, the DST is used for industrial applications such as a body armor and a traction control.

The DST has attracted the attention of physicists~\cite{Barnes89,Mewis11,Brown14,Lootens05,Cwalina14} as a typical nonequilibrium discontinuous phase transition between a liquid-like phase and a solid-like phase.
Apart from other important factors~\cite{Brown09,Brown10,Waitukaitis12}, it has been recognized recently that the mutual friction between grains plays an important role in the DST for dense suspensions~\cite{Otsuki11,Seto13,Mari14,Bi11,Pica11,Haussinger13,Kawasaki14}.
In addition, the normal stress difference becomes large when the shear thickening takes place~\cite{Lootens05,Cwalina14}.
The mechanism of the DST can also be understood by the introduction of an order parameter which exhibits a S-shape in
a plane of stress-strain rate (flow curve)~\cite{Wyart14,Nakanishi11,Nakanishi12,Nagahiro13,Grob14}.

Although most of previous studies on shear thickening are oriented to dense suspensions,
it would be convenient to consider relatively low density systems where kinetic theory tools ~\cite{Brilliantov04,Brey98,Garzo99,Lutsko05,Garzo13} can provide
a deeper understanding on the microscopic mechanisms involved in the DST.
Indeed, some papers have reported that a DST-like process for the kinetic temperature can take place as a result of a saddle-node bifurcation~\cite{Tsao95,BGK2016,DST16,Saha17}.
Thus, Tsao and Koch~\cite{Tsao95}
demonstrated the existence of a non-equilibrium discontinuous phase transition for the kinetic temperature between
a quenched state (a low temperature state) and an ignited state (a high temperature state) in a simple shear flow of a
(granular) gas-solid suspension described by the Boltzmann kinetic equation.
Recently, other works ~\cite{BGK2016,DST16,Saha17} have identified the discontinuous quenched-ignited transition with the DST if the system is agitated by thermal fluctuations. The validity of these studies have been verified from the event-driven Langevin simulation for hard spheres (EDLSHS)~\cite{Scala12} and the direct simulation Monte Carlo method \cite{B94}.
Such gas-solid suspensions are usually discussed in the context of fluidized beds~\cite{Gidaspow94,Jackson00} which
might be categorized as one of the typical inertial suspensions~\cite{Koch01}.
In particular, the homogeneous phase achieved by the balance between the injected gas flow from the bottom of a container and the gravity in fluidized beds is the target of our study.
It is remarkable that the previous studies on dilute gas-solid suspensions suggested that the DST
(or the discontinuous quenched-ignited transition) tends towards the CST (or the continuous quenched-ignited transition)
as the density increases~\cite{BGK2016,Saha17}.
Notice that the Newtonian branch for low shear rates disappears if the thermal agitation is absent.
 As a result, one can only observe the CST in the rheology for such systems~\cite{Tsao95,Chamorro15},
though the discontinuous ignited-quench transition can still be observed for the kinetic temperature.
These results are consistent with the analysis made by Santos {\it et al.} \cite{Santos98} which found the existence of a CST in moderately dense hard-core gases by using the revised Enskog theory.

It is worth noting that most of the previous theoretical studies of the above solid-gas suspensions~\cite{Tsao95,BGK2016,DST16,Chamorro15} are based on the application of
Grad's moment method~\cite{Grad49} to the Boltzmann~\cite{Garzo02,Santos04} and Enskog ~\cite{Garzo13} kinetic equations.
A slightly different method has been recently adopted by Refs.~\cite{Saha17,Lutsko04,Saha14,Saha16} since they consider an anisotropic Maxwellian distribution which reduces to the Maxwellian in the isotropic limit.
Although the latter solution can be more appropriate for highly dissipative sheared suspensions,
 it is quite intricate and requires some additional approximations to get explicit results.
In this context, the conventional Grad's moment method (which is based on the assumption that the distribution function is a local Maxwellian times a sum over Hermite polynomials) is simple enough to reproduce for instance the normal stress differences~\cite{Lutsko04,Saha14,Saha16}. Therefore, the conventional Grad's moment method can be still considered as a powerful method to describe the rheology of gas-solid suspensions.

Although the previous achievements of Refs.~\cite{Tsao95,DST16,Saha17} are remarkable,
they are limited to the low-density regime and hence their predictions are far from typical experimental situations.
 One of the few works devoted to dense gases was carried out by Sangani {\it et al.} \cite{Sangani96} two decades ago.
 In this paper, the authors extended the analysis of Ref.~\cite{Tsao95} to moderate densities by considering the Enskog equation.
Their analysis showed that the discontinuous transition of the kinetic temperature for dilute suspensions becomes continuous at relatively low density~\cite{Sangani96}.
 This conclusion agrees with the previous theories ~\cite{BGK2016,Saha17} for dilute suspensions.
However, the treatment of Sangani {\it et al.}~\cite{Sangani96} is not completely systematic since they ignore
the effects of thermal fluctuations.

The purpose of this paper is to extend the previous dilute results to moderately dense systems by solving the Enskog kinetic equation~\cite{Garzo99,Lutsko05,Garzo13,Resibois77} by two complementary and independent routes:
(i) Grad's moment method and (ii) event-driven simulations (EDLSHS).
The influence of the background fluid on particles is modeled via an external force constituted by two terms:
(i) a viscous drag force which mimics the friction of solid particles with the interstitial fluid and
 (ii) a stochastic Langevin-like term accounting for thermal fluctuations.  To assess the finite density effects on rheology, a set of coupled equations of the stress tensor, the kinetic temperature and the anisotropic temperatures corresponding to the normal stress differences are derived from Grad's approximation.
The validity of  our simple theory is also examined through a comparison with computer simulations.
The motivation of the the present work is twofold. First, since there is some evidence~\cite{Chialvo13} that the Enskog theory is accurate for solid volume fractions smaller than 0.5, our results will allow us to analyze the behavior of rheology for moderately dense suspensions corresponding to typical experiments.
 As a second point, our results will allow us to clarify whether the scenario proposed by Sangani {\it et al.}~\cite{Sangani96} is universal.

The organization of this paper is as follows.
The outline of the Enskog kinetic theory of moderately dense suspensions under a simple shear flow is briefly summarized in Sec.\ \ref{Enskogbis}. Section \ref{rheology} discusses the rheology of the suspension model where the details of the calculations appear in a series of Appendices.
Theoretical results are compared against computer simulations in Sec.\ \ref{simulation} for two values of the restitution coefficient $e$ ($e=1$ and $0.9$) and several values of the solid volume fraction $\varphi$ in the main text. 
As a complement, to assess the influence of inelasticity on rheology, theory and simulation results are also displayed in the Appendix G for the density $\varphi=0.3$ and several values of the restitution coefficient ($e=1,0.9,0.7,0.5$, and $0.3$).
 Section \ref{DST} deals with the transition from DST to CST. The paper is closed in Sec.\ \ref{discussion} where the results reported here are briefly discussed.

\section{Enskog kinetic equation for suspensions under simple shear flow}
\label{Enskogbis}

\subsection{Enskog kinetic equation for sheared granular suspensions}

Let us consider a collection of monodisperse smooth spherical grains of diameter
$\sigma$, mass $m$, and restitution coefficient $e$ satisfying $0< e \le 1$.
Because we are interested in the homogeneous state of fluidized beds,
the solid particles are distributed in a $d-$dimensional space only influenced by the background fluid under a uniform shear flow.
 This state is macroscopically characterized by a constant number density $n$, a uniform kinetic temperature $T$,
and macroscopic velocity field $\bm{u}=(u_x,\bm{u}_\perp)$, where the constant shear rate $\dot\gamma$ is given by
\begin{equation}
\label{plane_shear}
u_x=\dot\gamma y, \quad \bm{u}_\perp=\bm{0} .
\end{equation}
Let us introduce the peculiar momentum of $i-$th particle as $\bm{p}_i\equiv m(\bm{v}_{i}-\dot\gamma y \bm{e}_x)$, where $\bm{v}_i$ is the instantaneous velocity of $i-$th particle, and $\bm{e}_x$ is the unit vector parallel to $x$ direction. For low Reynolds numbers, a reliable model for describing solid particles immersed in a fluid (suspensions) is the Langevin equation
\begin{equation}
\label{Langevin_eq}
\frac{d{\bm{p}}_i}{dt}=-\zeta \bm{p}_i + \bm{F}_i^{({\rm imp})}+ m\bm{\xi}_i,
\end{equation}
where we have assumed that the solid particles are suspended by the gas flow and the gravity does not play any role.
We have also introduced the impulsive force $\bm{F}_i^{(\rm imp)}$ to express collisions between grains and the noise $\bm{\xi}_i(t)=\xi_{i,\alpha}(t)\bm{e}_\alpha$ has the average properties
\begin{equation}
\label{noise}
\langle \bm{\xi}_i(t)\rangle=0, \quad
\langle \xi_{i,\alpha}(t)\xi_{j,\beta}(t')\rangle = 2\zeta T_{\rm ex} \delta_{ij}\delta_{\alpha\beta}\delta(t-t').
\end{equation}
Here, the parameters $\zeta$ and $T_{\rm ex}$ characterize the drag from the background fluid and the
environmental temperature, respectively. Actually, the drag coefficient $\zeta$ should be a resistance matrix as a result of the hydrodynamic interactions between grains which strongly depends on the configuration of grains. For simplicity, however, we regard $\zeta$ as a scalar function of the average volume fraction $\varphi$ defined as
\begin{equation}
\label{volume_fraction}
\varphi=\frac{\pi^{d/2}}{2^{d-1} d \Gamma\left(\frac{d}{2}\right)}
n\sigma^d,
\end{equation}
where $\Gamma(x)=\int_0^\infty dt e^{-t}t^{x-1}$ is the Gamma function. This is a mean field approximation where the drag coefficient $\zeta$ is independent of the configuration of grains.
This simple model might be applicable to the description of inertial suspensions in which the mean diameter of suspended particles is approximately ranged from 1$\mu$m to 70$\mu$m~\cite{Koch01}.
In this paper, we assume that $\zeta \propto \eta_0 \propto \sqrt{T_{\rm ex}}$, where $\eta_0$ is the viscosity of the solvent or fluid phase.
If we ignore the density dependence of $\zeta$ and the grains are bidisperse soft spheres,
the Langevin model~\eqref{Langevin_eq} is equivalent to that used by Kawasaki {\it et al}.~\cite{Kawasaki14}.

So far, we did not specify the explicit dependence of $\zeta$ on $\varphi$ and $T_{\rm ex}$.
Let us rewrite $\zeta$ as
\begin{equation}
\label{zeta=zeta_0*R}
\zeta=\zeta_0 R(\varphi) ,
\end{equation}
where $\zeta_0=3\pi \eta_0 \sigma\propto \sqrt{m T_{\rm ex}}/\sigma$ and the solvent viscosity $\eta_0\propto \sqrt{m T_{\rm ex}}/\sigma^2$ for $d=3$. We adopt the following empirical expressions for the dimensionless resistance $R(\varphi)$:
\begin{equation}
R(\varphi)=
1+3\displaystyle\sqrt{\frac{\varphi}{2}}
\end{equation}
for $\varphi\le 0.1$~\cite{Tsao95,Koch90},
and
\begin{equation}
\label{Sangani3_18}
R(\varphi)= k_1(\varphi)-\varphi g_0(\varphi) \ln \epsilon_m
\end{equation}
for $\varphi>0.1$~\cite{Sangani96}. 
Here, $g_0(|\bm{r}|=\sigma,\varphi)$ is the radial distribution at contact,
which is believed to be uniform in the simple shear flow problem. 
For hard spheres ($d=3$) and $\varphi<0.49$, a good approximation for the radial distribution is~\cite{CS}
\begin{equation}\label{radial_fn}
g_0(|\bm{r}|=\sigma,\varphi)=\frac{1-\varphi/2}{(1-\varphi)^3}.
\end{equation}
Hereafter, we will use $g_0\equiv g_0(|\bm{r}|=\sigma,\varphi)$ as the abbreviation. 
In addition, in Eq.\ \eqref{Sangani3_18},
$\epsilon_m$ is the gap parameter characterizing the lubrication force between rough spheres, and $k_1(\varphi)$ for $d=3$ is the empirical function given by
\begin{equation}
k_1(\varphi)=1+\frac{3}{\sqrt{2}}\varphi^{1/2}+\frac{135}{64}\varphi \ln\varphi+
11.26\varphi (1-5.1\varphi+16.57\varphi^2-21.77\varphi^3).
\end{equation}
Because $\epsilon_m$ is related to the limitation of continuum description of suspensions, it is difficult to present its microscopic expression. Nevertheless, it is known that typical values of $\epsilon_m$ are in the range 0.01-0.05.
In this paper we will take $\epsilon_m=0.01$ for the later explicit calculation following Ref.~\cite{Garzo2012}.


Let us assume now that the suspension is under simple shear flow.
At a microscopic level, the simple shear flow state is generated by Lees-Edwards boundary conditions~\cite{LE72} which are simply periodic boundary conditions in the local Lagrangian frame $\bm{V}=(v_x-\dot\gamma y)\bm{e}_x+\bm{v}_\perp$.
In this frame, the velocity distribution function $f(\bm{r},\bm{v},t)$ is uniform
\beq
\label{vic1}
f(\bm{r},\bm{v},t)=f(\bm{V},t),
\eeq
and the Enskog equation for the granular suspension becomes ~\cite{Hayakawa03,Chamorro15}
\begin{equation}
\label{Enskog}
\left(\frac{\partial}{\partial t}-\dot\gamma
V_{y}\frac{\partial}{\partial V_{x}}
\right)f(\bm{V},t)
=
\zeta\frac{\partial}{\partial \bm{V}} \cdot \left(
\left\{ \bm{V}+ \frac{T_{\rm ex}}{m} \frac{\partial}{\partial \bm{V}} \right\}
 f(\bm{V},t) \right)+
 J_\text{E}(\bm{V}|f,f).
\end{equation}
The Enskog collision operator $J_\text{E}[\bm{V}|f,f]$ is given by (See the Appendix \ref{Enskog_base} )
\begin{equation}
\label{J(V|f)}
J_{\text{E}}\left[{\bf V}_{1}|f,f\right] =\sigma^{d-1}g_0 \int d{\bf V}
_{2}\int d\widehat{\boldsymbol{\sigma}}\,\Theta (\widehat{{\boldsymbol {\sigma}}}
\cdot \bm{v}_{12})(\widehat{\boldsymbol {\sigma }}\cdot \bm{v}_{12})\left[
\frac{
f(\bm{V}_1'',t)
f(\bm{V}_2''+\dot\gamma\sigma \widehat{\sigma}_y \bm{e}_x,t)}{e^2}
-f(\bm{V}_1,t)f(\bm{V}_2-\dot\gamma\sigma \widehat{\sigma}_y \bm{e}_x,t)
\right].
\end{equation}
In Eq.\ \eqref{J(V|f)}, the Heaviside step function is defined as $\Theta(x)=1$ for $x\ge 0$ and $\Theta(x)=0$ otherwise, $\bm{v}_{12}=\bm{V}_1-\bm{V}_2=\bm{v}_1-\bm{v}_2$ is the relative velocity at contact, and $\bm{\sigma}=\bm{r}_{12}$ where $\bm{r}_{12}\equiv \bm{r}_1-\bm{r}_2$.
In addition, the double primes in Eq.\ \eqref{J(V|f)} denote the pre-collisional velocities $\left\{\bm{V}_1^{''}, \bm{V}_2''\right\}$ that lead to $\left\{\bm{V}_1, \bm{V}_2\right\}$ following a binary collision:
\begin{equation}
\label{collision_rule}
\bm{V}_1^{''}=\bm{V}_1-\frac{1+e}{2e}(\bm{v}_{12}^{''}\cdot\widehat{\bm{\sigma}})\widehat{\bm{\sigma}}, \quad
\bm{V}_2^{''}=\bm{V}_2+\frac{1+e}{2e}(\bm{v}_{12}^{''}\cdot\widehat{\bm{\sigma}})\widehat{\bm{\sigma}}.
\end{equation}
In this paper we do not consider the effects of tangential friction and rotation induced by each binary collision.

The most important quantity in a shear flow problem is the pressure tensor $\sf{P}$. It has kinetic and collisional transfer contributions, i.e., $\sf{P}=\sf{P}^k+\sf{P}^c$. The kinetic contribution is
\begin{equation}
\label{pressure_tensor:kinetic}
P^k_{\alpha\beta}=m \int d\bm{V} V_\alpha V_\beta f(\bm{V}),
\end{equation}
while its collisional contribution is given by (see Appendix B for the derivation)
\begin{equation}
\label{pressure:collisonal}
P^c_{\alpha\beta}=\frac{(1+e)}{4} m\sigma^d g_0 \int d\bm{V}_1 \int d\bm{V}_2\int d\widehat{\bm{\sigma}}  \Theta(\bm{v}_{12}\cdot\widehat{\bm{\sigma}})
(\bm{v}_{12}\cdot\widehat{\bm{\sigma}})^2\widehat{\sigma}_\alpha\widehat{\sigma}_\beta
f\left(\bm{V}_1+\frac{1}{2}\dot\gamma \sigma \widehat{\sigma}_y \bm{e}_x\right)f\left(\bm{V}_2-\frac{1}{2}\dot\gamma \sigma \widehat{\sigma}_y \bm{e}_x\right).
\end{equation}
As usual, the hydrostatic pressure $P$ is defined as $P\equiv P_{\alpha\alpha}/d$,
where we adopt Einstein's rule for the summation {\it i.e.} $P_{\alpha\alpha}=\sum_{\alpha=1}^d P_{\alpha\alpha}$. The kinetic part of the pressure tensor satisfies the equation of the state of ideal gases, namely,  $P^k\equiv P^k_{\alpha\alpha}/d =n T$, where
\begin{equation}
\label{vic3}
n=\int d\bm{V} f(\bm{V})
\end{equation}
is the number density and
\begin{equation}
\label{kinetic_T}
T =\frac{1}{dn}\int d\bm{V} \bm{V}^2f(\bm{V})
\end{equation}
is the kinetic granular temperature.

\subsection{Grad's moment method}

The kinetic contribution $P^k_{\alpha\beta}$ to the pressure tensor can be obtained by multiplying both sides of Eq.\ \eqref{Enskog} by $m V_\alpha V_\beta$ and integrating over $\bm{V}$. The result is
\begin{equation}
\frac{\partial}{\partial t}P^k_{\alpha\beta}
+\dot\gamma (\delta_{\alpha x}P_{y \beta}^k+\delta_{\beta x} P_{y \alpha}^k)
=-2\zeta ( P_{\alpha\beta}^k- n T_{\rm ex} \delta_{\alpha\beta} )
-\Lambda_{\alpha\beta}^E ,
\label{Garzo31}
\end{equation}
where
\begin{equation}
\label{Garzo32}
\Lambda_{\alpha\beta}^E\equiv -m \int d\bm{V} V_\alpha V_\beta J_E(\bm{V}|f,f) .
\end{equation}
The collisional moment \eqref{Garzo32} can be rewritten as (see Appendix \ref{details_collision_transfer} for technical details)
\beq
\label{vic4}
\Lambda_{\alpha\beta}^E=\overline{\Lambda}_{\alpha\beta}^E+\dot\gamma (\delta_{\alpha x}P_{y \beta}^c+\delta_{\beta x}
P_{y \alpha}^c),
\eeq
where  $\overline{\Lambda}_{\alpha\beta}^E$ is defined by Eq.\ \eqref{over_Lambda} and we have omitted the last term on the right hand side of Eq.\ \eqref{total_Lambda_E} because we have accounted for that the heat flux vanishes in the simple shear flow problem by symmetry reasons [this can easily be deduced by considering Grad's distribution \eqref{Grad} as shown in the Appendix C.2]. 
Taking into account Eq.\ \eqref{vic4}, Eq.\ \eqref{Garzo31} reads
\begin{equation}
\label{Garzo31b}
\frac{\partial}{\partial t}P^k_{\alpha\beta}
+\dot\gamma (\delta_{\alpha x}P_{y \beta}+\delta_{\beta x} P_{y \alpha})
=-2\zeta ( P_{\alpha\beta}^k- n T_{\rm ex} \delta_{\alpha\beta} )
-\overline{\Lambda}_{\alpha\beta}^E.
\end{equation}


The simple shear flow state is in general non-Newtonian.
This can be characterized for instance by the anisotropic temperatures $\Delta T$ and $\delta T$ which are, respectively, defined as
\beq
\label{DT}
\Delta T \equiv \frac{P_{xx}^k-P_{yy}^k}{n},
\eeq
\beq
\label{dT}
\delta T \equiv \frac{P_{xx}^k-P_{zz}^k}{n}.
\eeq
Apart from the normal stresses, one can define a non-Newtonian shear viscosity coefficient $\eta(\dot\gamma,e)$ by
\beq
\label{shear_viscosity}
\eta(\dot\gamma,e)\equiv -\frac{P_{xy}}{\dot\gamma}.
\eeq

%
The time-dependent equations for $T$, $\Delta T$, $\delta T$, and $P_{xy}^k$ can be easily derived from Eq.\
\eqref{Garzo31b}.
They are given by
\begin{eqnarray}
\label{partT}
\frac{\partial}{\partial t} T&=&
-\frac{2\dot\gamma}{d n}P_{xy}+2\zeta (T_{\rm ex}-T) -
\frac{\overline{\Lambda}^E_{\alpha\alpha}}{d n},
\\
\label{part_DT}
\frac{\partial}{\partial t} \Delta T&=&
-\frac{2}{n}\dot\gamma P_{xy}-2\zeta \Delta T-\frac{\overline{\Lambda}_{xx}^E-\overline{\Lambda}_{yy}^E}{n},
\\
\label{part_dT}
\frac{\partial}{\partial t}\delta T&=&
-\frac{2}{n}\dot\gamma P_{xy}-2\zeta \delta T-\frac{\overline{\Lambda}_{xx}^E-\overline{\Lambda}_{zz}^E}{n},
\\
\label{part_P_{xy}}
\frac{\partial}{\partial t}P_{xy}^k&=&- \dot\gamma P_{yy}
-2\zeta P_{xy}^k-\overline{\Lambda}_{xy}^E.
\end{eqnarray}
The moment equations \eqref{partT}--\eqref{part_P_{xy}} are still exact and have been obtained without the explicit knowledge of the velocity distribution function $f$.

On the other hand, the exact expression of the collision integral $\overline{\Lambda}_{\alpha\beta}^E$ is not known, even in the elastic case. A good estimate of this collisional moment can be expected by using Grad's
approximation~\cite{Garzo13,BGK2016,Chamorro15,Grad49,Garzo02,Santos04}
\begin{equation}
\label{Grad}
f(\bm{V})=f_{\rm M}(\bm{V})\left(1+\frac{m}{2T}\Pi^k_{\alpha\beta}V_\alpha V_\beta
\right),
\end{equation}
where
\beq
\label{Maxwell}
f_{\rm M}(\bm{V})=
n\left(\frac{m}{2\pi T}\right)^{d/2}\exp\left(-\frac{m V^2}{2T} \right)
\eeq
is the Maxwellian distribution and
\beq
\label{vic5}
\Pi^k_{\alpha\beta}\equiv \frac{P^k_{\alpha\beta}}{nT}-\delta_{\alpha\beta}
\end{equation}
is the traceless part of the (dimensionless) kinetic pressure tensor $P^k_{\alpha\beta}$.
The collisional moment $\overline{\Lambda}_{\alpha\beta}^E$ can be determined when the trial distribution \eqref{Grad} is inserted into Eq.\ \eqref{over_Lambda}. After a lengthy algebra (see the Appendices \ref{details_collision_transfer} and \ref{derivation_Lambda} for details), one achieves the expression
\beqa
\label{Lambda_E:result}
\overline{\Lambda}^E_{\alpha\beta}&=&
g_0 nT \left\{
\nu \Pi^k_{\alpha\beta}+\lambda \delta_{\alpha\beta}
-\frac{2^{d-2}}{(d+2)(d+4)} \varphi  (1+e)\dot\gamma \left[(d+4)(1-3e)
(\delta_{\alpha x}\delta_{\beta y}+\delta_{\alpha y}\delta_{\beta x})\right.\right.
\nonumber\\
& & \left.\left.
+
2(d+1-3e)\left(\Pi^k_{\alpha x}\delta_{\beta y}+\Pi^k_{\alpha y}\delta_{\beta x}
+\Pi^k_{\beta x}\delta_{\alpha y}+\Pi^k_{\beta y}\delta_{\alpha x}\right)
-6(1+e)\delta_{\alpha \beta} \Pi^k_{xy}
\right]\right\}.
\end{eqnarray}
Here, the quantities $\nu$ and $\lambda$ are given, respectively, by~\cite{Garzo02,Santos04,DST16}
\beq
\label{nu}
\nu= \frac{\sqrt{2}\pi^{(d-1)/2}}{d(d+2)\Gamma\left(d/2\right)} (1+e)(2d+3-3e) n  \sigma^{d-1} v_T,
\eeq
\beq
\label{nu'}
\lambda=\frac{\sqrt{2}\pi^{(d-1)/2}}{d\Gamma\left( d/2\right)} (1-e^2)  n \sigma^{d-1}v_T ,
\eeq
where
$v_T=\sqrt{2T/m}$ is the thermal velocity.
Notice that upon deriving the expression \eqref{Lambda_E:result} for $\overline{\Lambda}^E_{\alpha\beta}$ nonlinear terms in $\Pi_{\alpha\beta}^k$ have been neglected. As will show below, for \emph{dilute} gases ($\varphi\to 0$),
this approximation yields $P_{xx}^k\neq P_{yy}^k$ but $P_{yy}^k=P_{zz}^k$.
The latter equality disagrees with computer simulation results \cite{Tsao95,Chamorro15}.
The evaluation of $\overline{\Lambda}^E_{\alpha\beta}$ for dilute gases by retaining all the quadratic terms in the pressure tensor has been carried out in Ref.\ \cite{Chamorro15}.
The inclusion of these nonlinear corrections allows us to determine the normal stress differences in the plane orthogonal to the shear flow (e.g., $P_{yy}-P_{zz}$).
Nevertheless, since this difference is small, the expression \eqref{Lambda_E:result} can be considered as accurate, even in the limit of dilute gases as demonstrated in Ref.\ \cite{DST16}.

The set of coupled differential equations \eqref{partT}--\eqref{part_P_{xy}} can be written more explicitly when one takes into account the result \eqref{Lambda_E:result}:
\begin{eqnarray}
\label{d_tT}
\frac{\partial}{\partial t} T
&=& -\frac{2\dot\gamma}{d n}{\cal C}_d P^k_{xy}-\frac{2\dot\gamma}{d n}P_{xy}^c
+2\zeta (T_{\rm ex}-T) -g_0 \lambda T, \\
\label{d_tDT}
\frac{\partial}{\partial t} \Delta T&=&-\frac{2}{n}\dot\gamma \left(P_{xy}^k+P_{xy}^c\right)-(\nu g_0+2\zeta) \Delta T
, \\
\label{d_tdT}
\frac{\partial}{\partial t}\delta T&=&
-\frac{2}{n}\dot\gamma \left({\cal E}_d P_{xy}^k+P_{xy}^c\right)-(\nu g_0+2\zeta)\delta T
,\\
\label{d_tP_{xy}}
\frac{\partial}{\partial t} P_{xy}^k
&=& \dot\gamma n\left(\frac{d-1}{d} {\cal D}_d \Delta T
-\frac{d-2}{d}{\cal E}_d \delta T-{\cal C}_d T\right)
-\dot\gamma P_{yy}^c
-(\nu g_0+2\zeta) P^k_{xy}.
\end{eqnarray}
Here, we have introduced the (dimensionless) quantities
\beq
\label{c_C}
{\cal C}_d(e,\varphi)=1-\frac{2^{d-2}}{d+2}(1+e)(1-3e)\varphi g_0,
\eeq
\beq
\label{c_E}
{\cal E}_d(e,\varphi)= 1-\frac{2^{d}}{(d+2)(d+4)}(1+e)(d+1-3e)\varphi g_0,
\eeq
\beq
\label{c_D}
{\cal D}_d(e,\varphi)= 1-\frac{2^{d-1}(d-2)}{(d-1)(d+2)(d+4)}(1+e)(d+1-3e)\varphi g_0.
\eeq
%
In addition, upon deriving Eqs.\ \eqref{d_tT}--\eqref{d_tP_{xy}} we have used the relations
\beq
\label{vic6}
\Pi_{xx}^k=\frac{\Delta T}{d T}+\frac{d-2}{d}\frac{\delta T}{T}, \quad
\Pi_{yy}^k=\frac{1-d}{d}\frac{\Delta T}{T}+\frac{d-2}{d}\frac{\delta T}{T}, \quad
\Pi_{zz}^k=\frac{\Delta T}{d T}-\frac{2}{d}\frac{\delta T}{T}.
\eeq

%
To close the problem, one still needs to compute the collisional transfer contributions $P_{\alpha\beta}^c$ to the pressure tensor.
This can be achieved by inserting Grad's distribution \eqref{Grad} into Eq.\ \eqref{pressure:collisonal}.
On the other hand, this computation yields an intricate expression for $P_{\alpha\beta}^c$ that must be numerically evaluated. 
Thus, in order to get simple and accurate results, only terms up to the first order in the shear rate are considered
in the above calculation. The final result is (see the Appendix \ref{collision_stress})
\beq
\label{P_c:main_text}
P^c_{\alpha\beta}\approx
2^{d-2}(1+e) \varphi g_0 n T
\left[
\delta_{\alpha\beta}+\frac{2}{d+2}\Pi^k_{\alpha\beta}
-\dot\gamma^* \tau_T \frac{2\sqrt{2}}{\sqrt{\pi}(d+2)}
\left(
\delta_{\alpha x}\delta_{\beta y}+\delta_{\alpha y}\delta_{\beta x}\right)
\right],
\eeq
where
\beq
\label{dimless_time}
\dot\gamma^*\equiv \frac{\dot\gamma}{\zeta_0}, \quad \tau_T=\frac{\zeta_0 \sigma}{v_T}.
\eeq
The quantity $\zeta_0$ is defined in Eq.\ \eqref{zeta=zeta_0*R}.
Since $\zeta_0 \propto \sqrt{T_\text{ex}}$ and $v_T \propto \sqrt{T}$, the parameter $\tau_T$ measures the competing effect between
the environmental temperature $T_\text{ex}$ and the kinetic temperature $T$.
In the case that the environmental temperature $T_{\rm ex}$ is much lower than the kinetic temperature, then $\tau_T$ can be considered as a small parameter and could be neglected in the expression \eqref{P_c:main_text} of the collision contribution to the pressure tensor. In fact, as we will show below, the theoretical predictions compare better with simulations when we neglect this term ($\tau_T=0$) in Eq.\ \eqref{P_c:main_text}. In this context, one could argue for that the results derived here could be relevant for situations where the stresses applied by the background fluid on solid particles have a weak influence on the dynamics of grains.

It is important to remark that the use of the expression \eqref{P_c:main_text} is mainly motivated by the desire of analytic expressions for the rheological properties that allow to unveil in a clean way the impact of both the restitution coefficient $e$ and the (scaled) shear rate $\dot\gamma^*$ on momentum transport.
Of course, since the collisional transfer contribution $P_{\alpha\beta}^c$ are expected to strongly depend
on $\dot\gamma^*$ in the steady state~\cite{Santos04}, the truncation made in Eq.\ \eqref{P_c:main_text} can be likely only justified for nearly elastic systems.
On the other hand, as we will show in Sec.\ IV, the good agreement found between theory and simulations for moderately strong dissipation
(i.e., $e=0.9$) justifies the use of the expression \eqref{P_c:main_text} beyond the elastic limit ($e \to 1$).

%
After a transient period one expects that the system achieves a steady state. In this steady state, the viscous heating term ($-\dot\gamma P_{xy}>0$) is exactly balanced by the cooling terms arising from the collisional dissipation and the friction between the background fluid and the solid particles.
One of the main goals of this paper is to determine the rheological properties of the gas-solid suspension in the steady state.
This will be carried out analytically in the next section by solving the set of coupled equations \eqref{partT}--\eqref{part_P_{xy}} when $\partial_t\to 0$.


\section{Rheology for steady simple shear flow}
\label{rheology}


%
As mentioned before, the rheology of gas-solid suspensions are determined in this section by solving the constitutive
equations \eqref{d_tT}-\eqref{d_tP_{xy}} in the steady state. First, in order to solve this set of equations, it is convenient to write it in dimensionless form. To do that, since $\zeta\propto \sqrt{T_{\rm ex}}R(\varphi)$, we introduce here the reduced quantities
\begin{equation}
\label{nu**}
\nu^{*}=\frac{\nu}{\sqrt{\theta}\zeta_0R(\varphi)}, \quad
{\lambda}^{*}=\frac{{\lambda}}{\sqrt{\theta}\zeta_0R(\varphi)},
\end{equation}
where $\theta\equiv T/T_{\rm ex}$.
In terms of the above quantities, in the steady state, Eqs.\ \eqref{d_tT}-\eqref{d_tP_{xy}} read
\beq
\label{steady1}
-\frac{2\dot \gamma^*}{d R}\left(\mathcal{C}_d \Pi_{xy}^k+P_{xy}^{c*}\right)=g_0\sqrt{\theta}\lambda^*+2(1-\theta^{-1}),
\eeq
\beq
\label{steady2}
-\frac{2\dot \gamma^*}{R}\left(\Pi_{xy}^k+P_{xy}^{c*}\right)=\left(2+g_0\sqrt{\theta}\nu^*\right)\frac{\Delta \theta}{\theta},
\eeq
\beq
\label{steady3}
-\frac{2\dot \gamma^*}{R}\left(\mathcal{E}_d\Pi_{xy}^k+P_{xy}^{c*}\right)=\left(2+g_0\sqrt{\theta}\nu^*\right)\frac{\delta \theta}{\theta},
\eeq
\beq
\label{steady4}
\frac{\dot\gamma^*}{R}P_{yy}^{c*}
+\left(2+\nu^*g_0 \sqrt{\theta}\right)\Pi_{xy}^k=-
\frac{\dot\gamma^*}{R} \left(\frac{d-1}{d}
{\cal D}_d\frac{\Delta \theta}{\theta}-\frac{d-2}{d}{\cal E}_d\frac{\delta \theta}{\theta}
-{\cal C}_d\right),
\eeq
where $P_{ij}^{c*}\equiv P_{ij}^c/nT$, $\Delta \theta\equiv \Delta T/T_\text{ex}$ and $\delta \theta\equiv \Delta T/T_\text{ex}$.
The solution to Eqs.\ \eqref{steady1}--\eqref{steady3} can be written as
\beq
\label{steady5}
\Pi_{xy}^k=\frac{d R\left[2(1-\theta)-g_0 \theta^{3/2}\lambda^*\right]+2\sqrt{\frac{2}{\pi}}{\cal F}_d \tau_T\theta \dot\gamma^{*2}}
{2\left({\cal C}_d+{\cal F}_d\right) \theta \dot\gamma^*},
\eeq
\beq
\label{steady6}
\frac{\Delta \theta}{\theta}=\frac{2\sqrt{\frac{2}{\pi}}{\cal F}_d \left({\cal C}_d-1\right)\tau_T\theta \dot\gamma^{*2}+d\left(1+{\cal F}_d\right)R\left[2(\theta-1)+g_0\theta^{3/2}\lambda^*\right]}{R\left({\cal C}_d+{\cal F}_d\right)(2+g_0 \sqrt{\theta}\dot\gamma^*)},
\eeq
\beq
\label{steady7}
\frac{\delta \theta}{\theta}=\frac{2\sqrt{\frac{2}{\pi}}{\cal F}_d\left({\cal C}_d-{\cal E}_d\right)\tau_T\theta \dot\gamma^{*2}+d\left({\cal E}_d+{\cal F}_d\right)R\left[2(\theta-1)+g_0\theta^{3/2}\lambda^*\right]}{R\left({\cal C}_d+{\cal F}_d\right)(2+g_0 \sqrt{\theta}\dot\gamma^*)},
\eeq
where
\beq
\label{Fd}
{\cal F}_d=\frac{2^{d-1}(1+e)}{d+2}g_0 \varphi.
\eeq
Upon deriving Eqs.\ \eqref{steady5}--\eqref{steady7}, use has been made of Eq.\ \eqref{P_c:main_text} for the collision transfer contribution to the pressure tensor. Finally, when Eqs.\ \eqref{steady5}--\eqref{steady7} are substituted into Eq.\ \eqref{steady4}, one achieves the following quartic equation in $\dot\gamma^*$:
\begin{equation}
\label{quartic}
R^4\mathscr{C}_4(e,\varphi,\theta) \dot\gamma^{*4}+R^2
\mathscr{C}_2(e,\varphi,\theta) \dot\gamma^{*2}+
\mathscr{C}_0(e,\varphi,\theta)=0.
\end{equation}
The coefficients $\mathscr{C}_4$, $\mathscr{C}_2$, and $\mathscr{C}_0$ are nonlinear functions of the restitution coefficient $e$, the volume fraction $\varphi$, and the (scaled) kinetic temperature $\theta$. Their explicit forms are given in the Appendix \ref{vicente}.

Although an explicit expression of $\theta$ in terms of $e$, $\varphi$, and $\dot\gamma^*$ is not known, the dependence of $\theta$ on the latter parameters can be implicitly obtained from the physical solution to Eq.\ \eqref{quartic} as $\dot\gamma^{*2}(\theta,e,\varphi)$. Once $\theta$ is known, the remaining rheological functions can be determined from Eqs.\ \eqref{steady5}--\eqref{steady7} in terms of $e$, $\varphi$, and $\dot\gamma^*$. In the low-density limit ($\varphi\to 0$), previous results \cite{DST16} obtained for dilute granular suspensions are recovered.

On the other hand, given that the collisional stress has been obtained by retaining terms up to the first order in the shear rate, for practical purposes it is more convenient to consider the limit $\tau_T \to 0$ but finite $e$ and $\varphi$ in the quartic equation \eqref{quartic}. In this case, we can write
\beq
\label{quartic.2}
\dot\gamma^*=\dot\gamma_0+\dot\gamma_1 \tau_T+\cdots,
\eeq
where the coefficients $\dot\gamma_0$ and $\dot\gamma_1$ can be easily obtained from the quartic equation \eqref{quartic} as
\beq
\label{quartic.3}
\dot\gamma_0
=\frac{1}{R}\sqrt{-\frac{\mathscr{C}_0}{\mathscr{C}_2^{(0)}}},
\eeq
\beq
\label{quartic.4}
\dot{\gamma}_1=-\frac{\mathscr{C}_2^{(1)}\dot\gamma_0+\mathscr{C}_4^{(0)}R^2 \dot\gamma_0^3}{2\mathscr{C}_2^{(0)}}.
\eeq
The quantities $\mathscr{C}_4^{(0)}$, $\mathscr{C}_2^{(0)}$, and $\mathscr{C}_2^{(1)}$ are defined in the Appendix \ref{vicente}. As mentioned before, an accurate and simple estimate of $\dot \gamma^*$ is provided by its zeroth-order form $\gamma_0$.

In summary, for given values of the restitution coefficient and density, Eq.\ \eqref{quartic.2} gives the shear-rate dependence of the (scaled) kinetic temperature $\theta$. The stress tensor $P_{xy}^*\equiv P_{xy}/nT$ and the first $\Delta T$ and second $\delta T$ stress normal differences are obtained by substituting $\theta(\dot\gamma^*)$ into Eqs.\ \eqref{steady1}--\eqref{steady3}, respectively.
The reliability of these theoretical results will be assessed in Sec.\ \ref{simulation} via a comparison against computer simulations.

\section{Comparison between theory and simulation}
\label{simulation}

\begin{figure}[htbp]
\includegraphics[width=140mm]{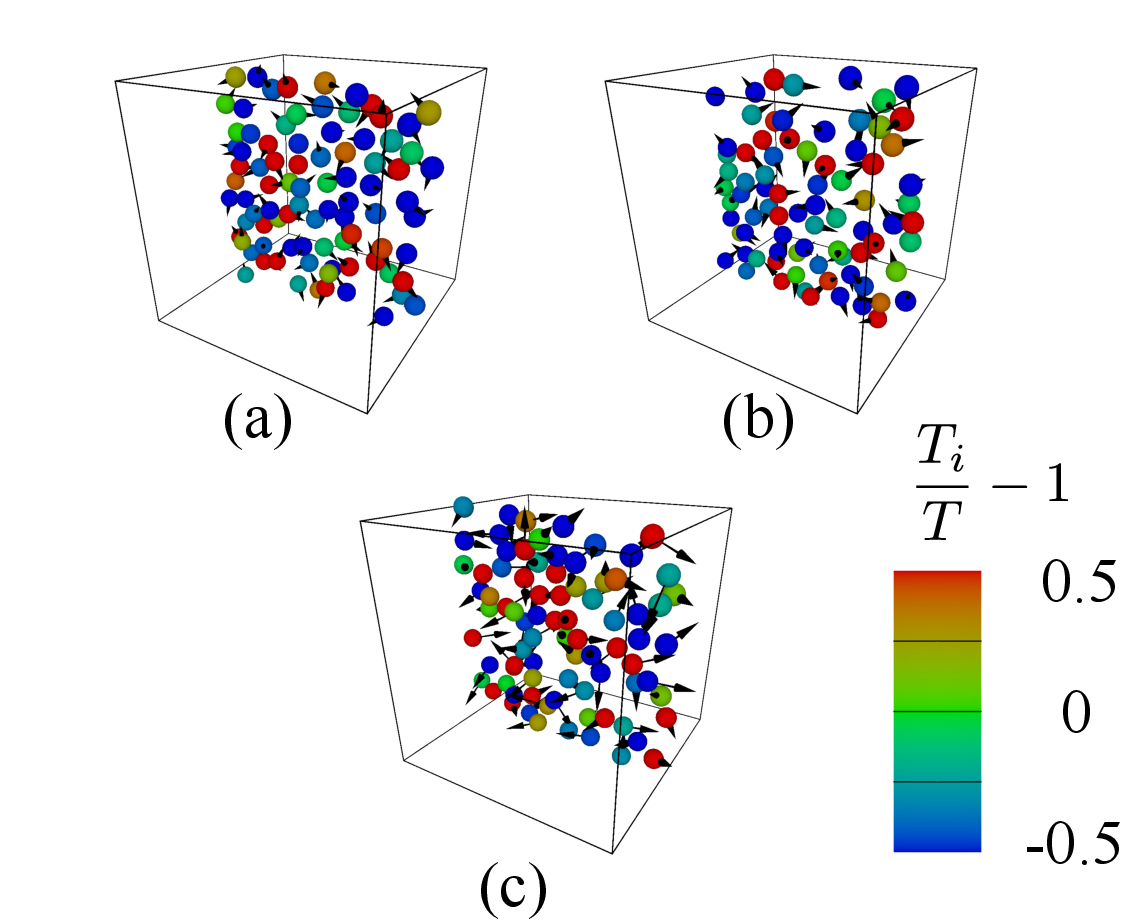}
\caption{
(Color online)
Plots of the configuration of particles and
the displacement vectors (black arrows) during the interval $1.0/\zeta_0$.
in cross sections for the shear rates (a) $\dot\gamma^*=1.0$ , (b) $3.0$, and (c) $10.0$.  The restitution coefficient is $e = 0.90$ while the density is $\varphi=0.3$. Notice that the uniform shear term is subtracted in the displacement vector.
We also show the temperature for the $i$-th particle $T_i \equiv (1/N) \sum_{i=1}^N m(\bm{v}_i - \bm{u})2/d$.
The color indicates the magnitude of $T_i/T - 1$.
  }
\label{fig_snapshot}
\end{figure}


%
The goal of this section is to validate our theoretical results by using the EDLSHS. 
We consider Lees-Edwards boundary conditions in a three-dimensional ($d=3$) periodic box~\cite{LE72,Scala12}.
Under these conditions, the Langevin equation \eqref{Langevin_eq} is equivalent to
Eqs.~\eqref{Enskog} and \eqref{J(V|f)_app}, when molecular chaos ansatz is assumed.
Therefore, if we can approximate Eq.~\eqref{J(V|f)_app} by the Enskog collision operator \eqref{J(V|f)}, our theory gives a good approximation of Eq.~\eqref{Langevin_eq}.

Notice that it is difficult to adopt neither the conventional event-driven simulation nor the soft-core simulation for our problem. The existence of both 
the inertia term $d\bm{p}/dt$ and the drag term proportional to $\zeta$ in Eq.~\eqref{Langevin_eq} makes difficult the use of the conventional event-driven simulation. In addition, a sudden increment of the viscosity in the vicinity of a DST gives rise to numerical difficulties of soft-core simulations.
Thus, to avoid the above difficulties, we adopt in this paper the EDLSHS. This is in fact a powerful simulator for hard spheres under the influence of the drag and the inertia terms with the aid of Trotter decomposition~\cite{Scala12}
 (some details of the EDLSHS method are provided in the Appendix \ref{EDLSHS}).

In our simulations, we fix the number of grains $N=1000$ as well as the background fluid temperature $T_{\rm ex}^*\equiv T_{\rm ex}/(m\sigma^2\zeta_0^2)=0.01$. 
Several volume fractions $\varphi$ are considered: $\varphi=0.01, 0.05, 0.10, 0.20, 0.30, 0,40$ and 0.50. The first density corresponds to a dilute suspension while the latter can be considered as a relatively high dense suspension.
Notice that previous works \cite{LBD02,DHGD02,MGAL06,MDCPH11,MGH14} have shown that the results derived from
the Enskog equation are quite accurate for moderately dense systems (for instance, $\varphi \lesssim 0.2$ for $d=3$).
Two different values of the restitution coefficient $e$ are considered in this section: $e=1$ (elastic grains) and $e=0.9$ (granular grains with moderate inelasticity) in the main text. 
More inelastic systems are considered in the Appendix \ref{echange} for the density $\varphi=0.3$.
All the rheological variables presented in this paper are measured after the system reaches a steady state (for $t>400/\zeta_0$). In addition, all the variables are averaged by 10 ensemble averages which have different initial conditions and 10 time averages during the time intervals $10/\zeta_0$ for each initial condition.  We have confirmed that the fluctuations of the observables are sufficiently small.

Before considering the rheological properties of the gas-solid suspension, Fig.\ \ref{fig_snapshot} displays a snapshot of the configurations and displacements of particles in a cross section for each given set of parameters. In particular, the panels (a), (b) and (c) of Fig.~\ref{fig_snapshot} represent the quenched, intermediate and ignited states, respectively, for $e=0.9$ and $\varphi=0.3$.
Here, the intermediate state means the intermediate between the quenched and ignited states.
Only a configuration of particles in a cross section in each panel of Fig.~\ref{fig_snapshot} is displayed.
Because the motion and configuration of the moderately dense gas seem to be uniform, 
 the use of the (homogeneous) Enskog kinetic equation \eqref{Enskog} for describing the simple shear flow is justified.

\begin{figure}[htbp]
\includegraphics[width=150mm]{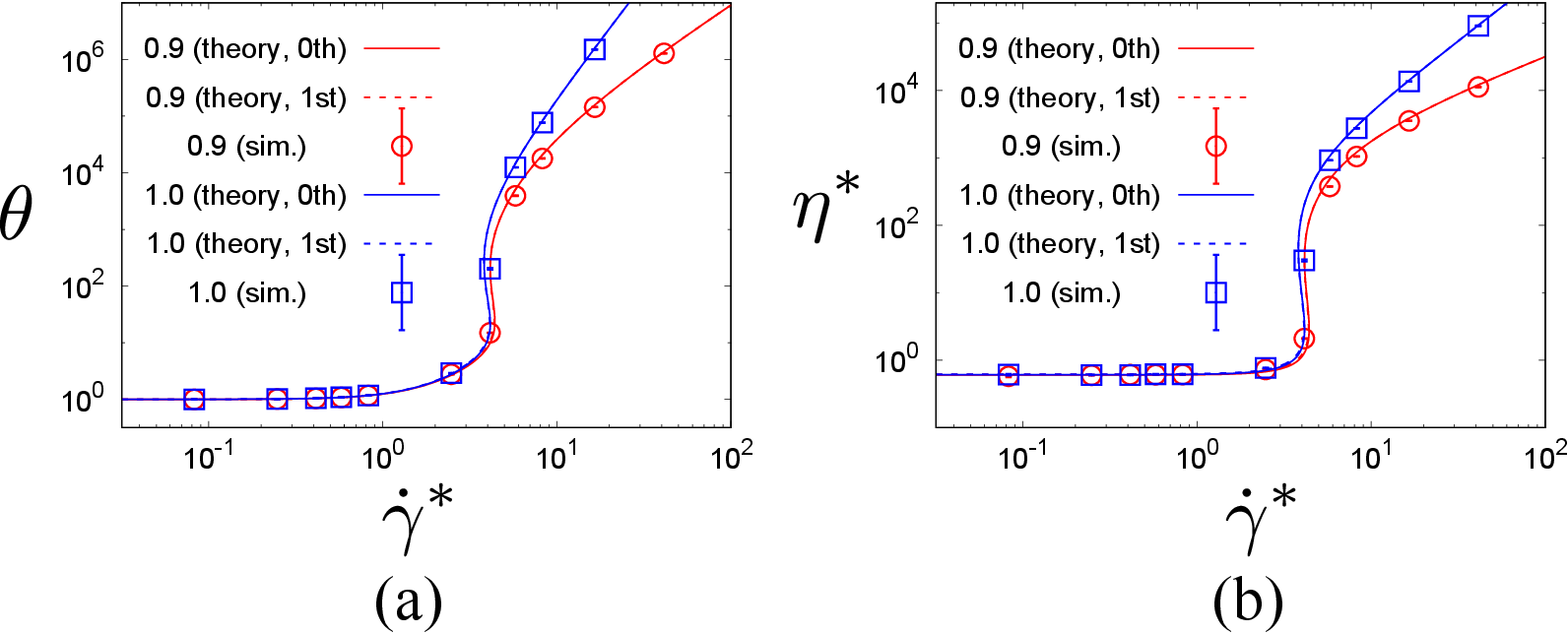}
\caption{
(Color online)
Plots of $\theta$ (panel (a)) and $\eta^*$ (panel (b)) versus the (scaled) shear rate $\dot\gamma^{*}$ for $\varphi=0.01$ and two different values of the restitution coefficient $e$: $e=1$ and $e=0.9$. The solid and dashed lines correspond to the (perturbative) theoretical results obtained in the zeroth-order (denoted by 0th in the legend) and first-order (denoted by 1st in the legend) in $\tau_T$, respectively. Symbols refer to computer simulation results.
  }
\label{fig1}
\end{figure}

\begin{figure}[htbp]
\includegraphics[width=150mm]{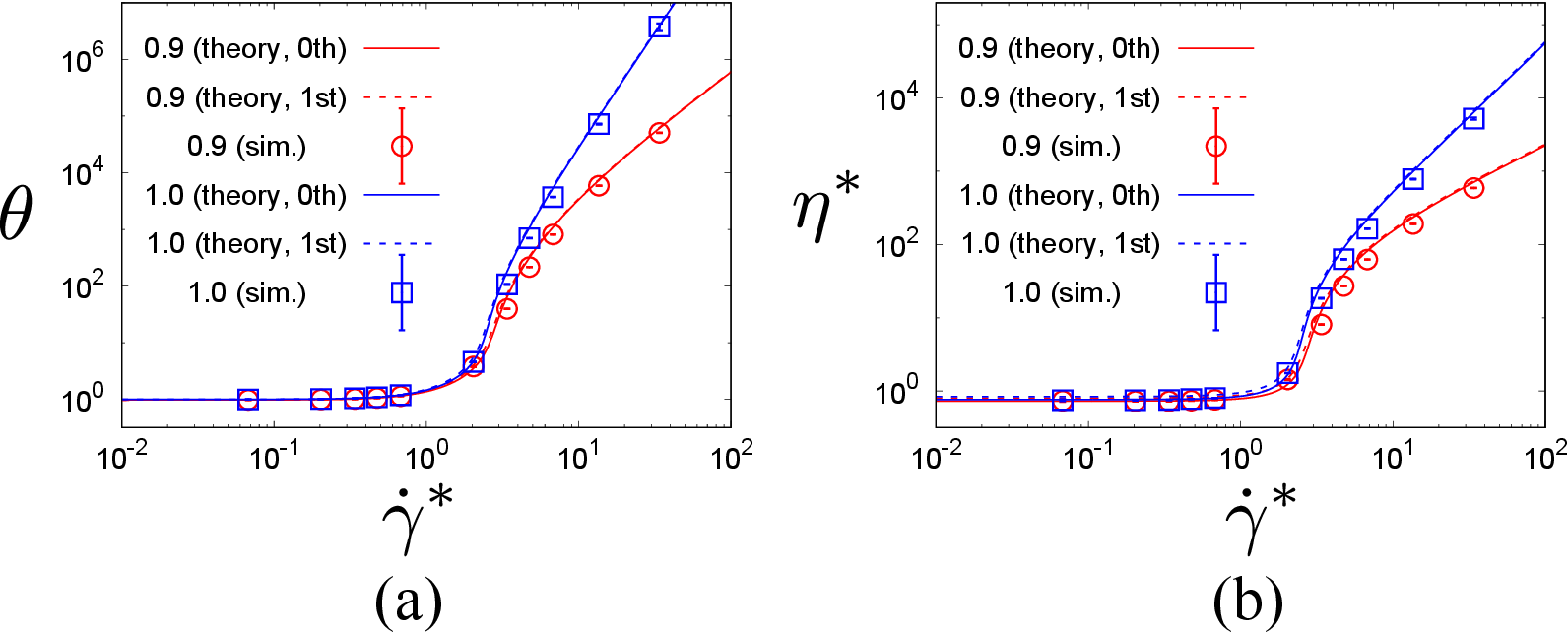}
\caption{
(Color online)
Plots of $\theta$ (panel (a)) and $\eta^*$ (panel (b)) versus the (scaled) shear rate $\dot\gamma^{*}$ for $\varphi=0.05$ and two different values of the restitution coefficient $e$: $e=1$ and $e=0.9$. The solid and dashed lines correspond to the (perturbative) theoretical results obtained in the zeroth-order (denoted by 0th in the legend) and first-order (denoted by 1st in the legend) in $\tau_T$, respectively. Symbols refer to computer simulation results.
  }
\label{fig2}
\end{figure}

\begin{figure}[htbp]
\includegraphics[width=150mm]{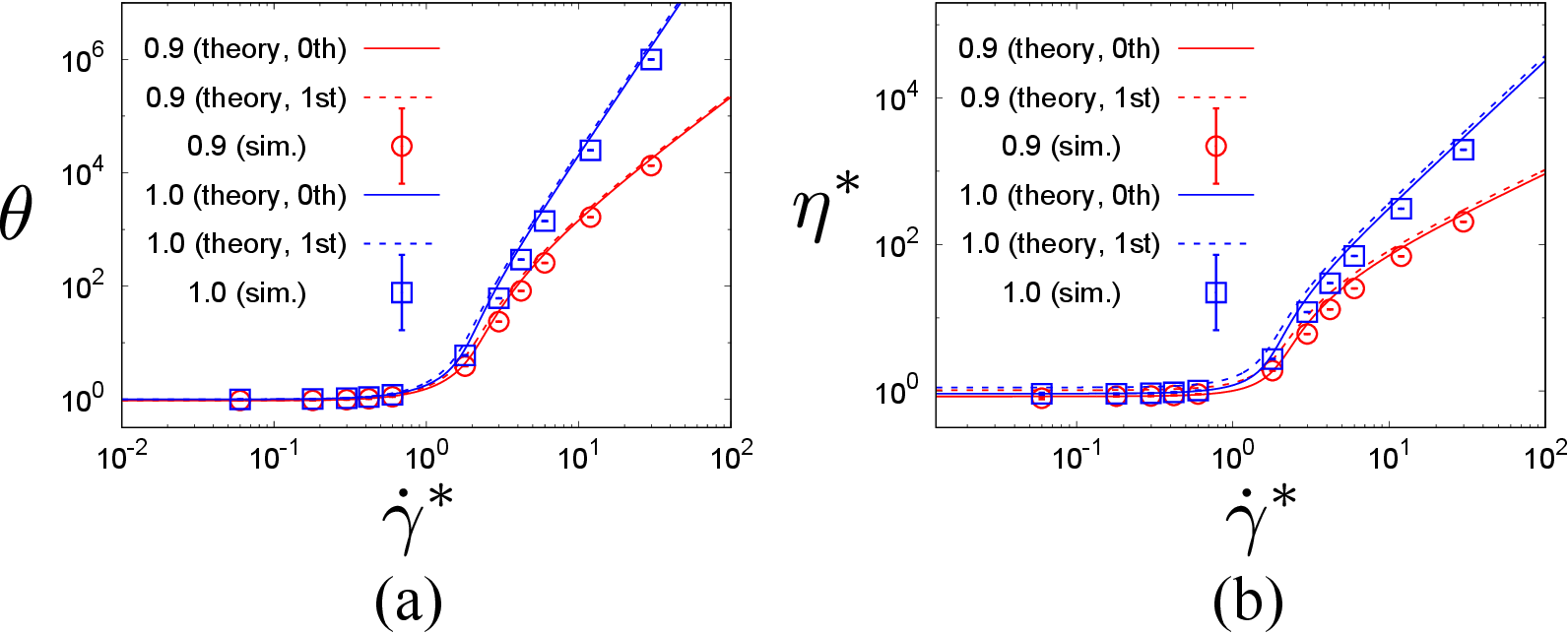}
\caption{
(Color online)
Plots of $\theta$ (panel (a)) and $\eta^*$ (panel (b)) versus the (scaled) shear rate $\dot\gamma^{*}$ for $\varphi=0.10$ and two different values of the restitution coefficient $e$: $e=1$ and $e=0.9$. The solid and dashed lines correspond to the (perturbative) theoretical results obtained in the zeroth-order (denoted by 0th in the legend) and first-order (denoted by 1st in the legend) in $\tau_T$, respectively. Symbols refer to computer simulation results.
  }
\label{fig3}
\end{figure}

\begin{figure}[htbp]
\includegraphics[width=150mm]{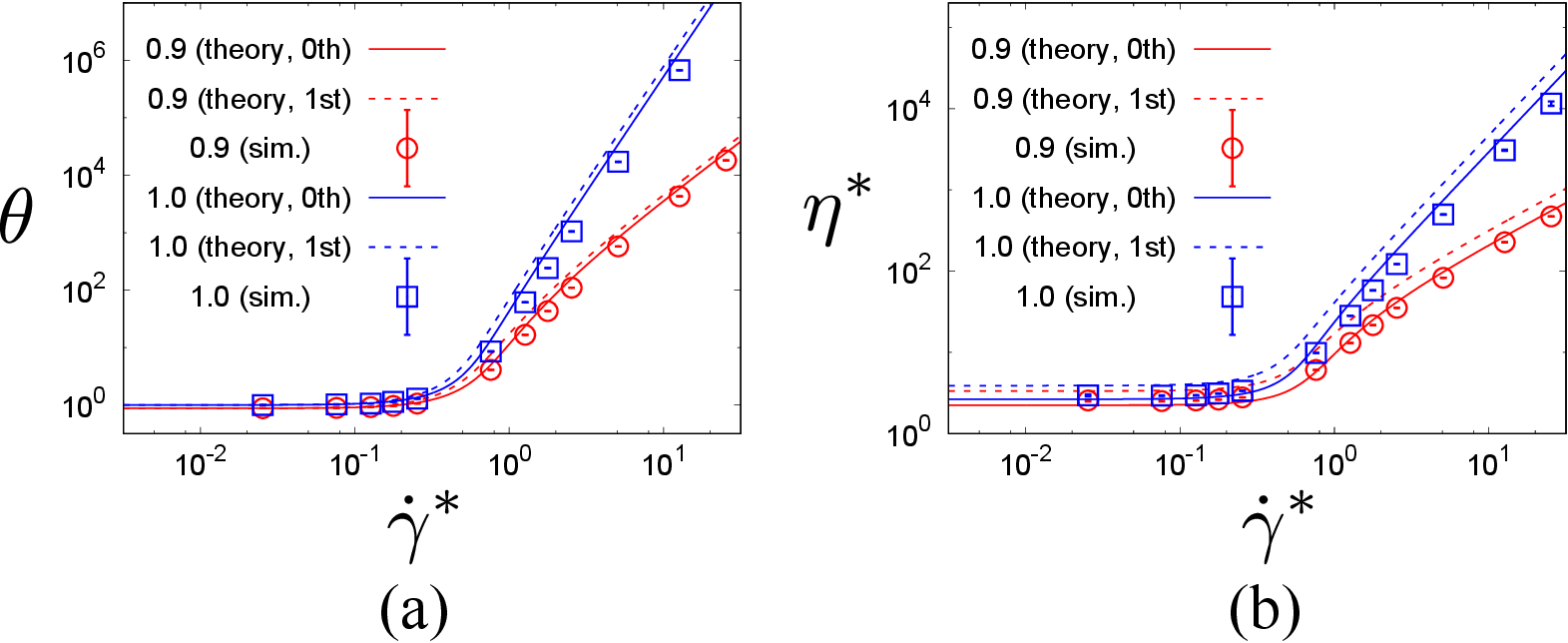}
\caption{
(Color online)
Plots of $\theta$ (panel (a)) and $\eta^*$ (panel (b)) versus the (scaled) shear rate $\dot\gamma^{*}$ for $\varphi=0.20$ and two different values of the restitution coefficient $e$: $e=1$ and $e=0.9$. The solid and dashed lines correspond to the (perturbative) theoretical results obtained in the zeroth-order (denoted by 0th in the legend) and first-order (denoted by 1st in the legend) in $\tau_T$, respectively. Symbols refer to computer simulation results.
  }
\label{fig4}
\end{figure}

\begin{figure}[htbp]
\includegraphics[width=150mm]{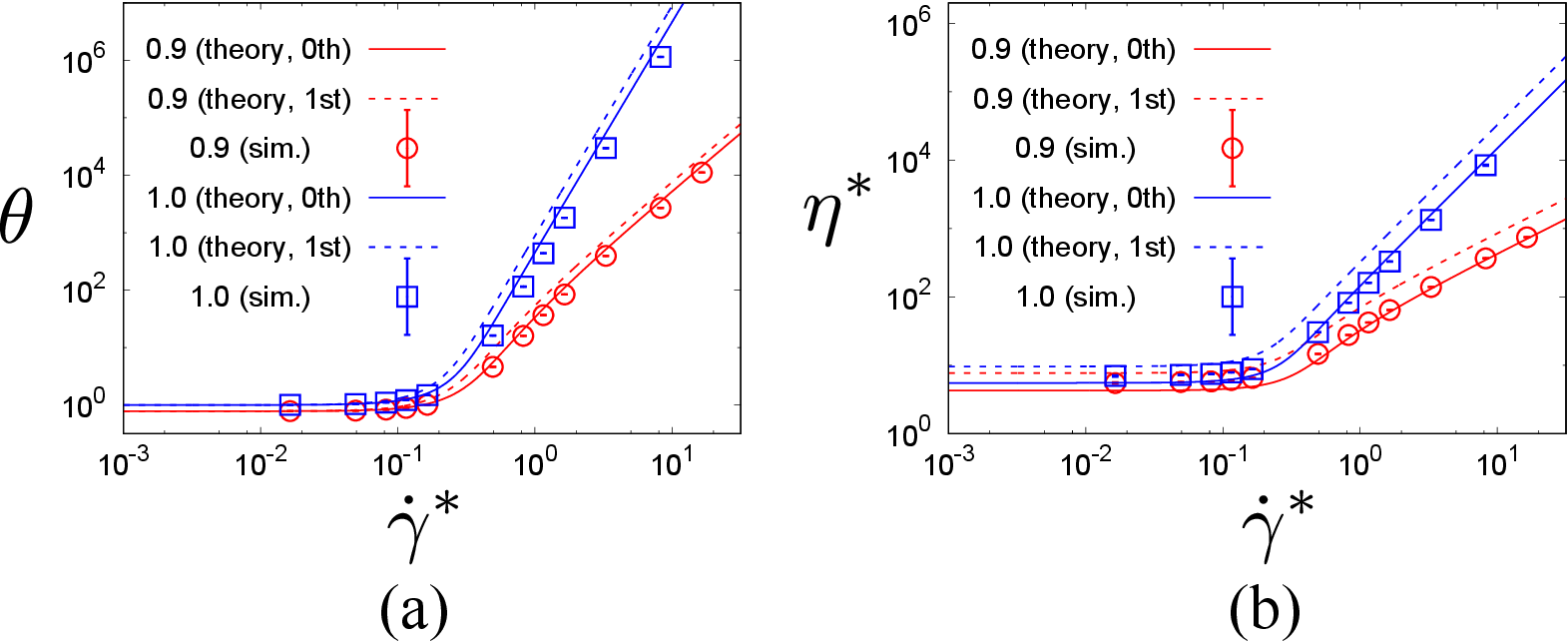}
\caption{
(Color online)
Plots of $\theta$ (panel (a)) and $\eta^*$ (panel (b)) versus the (scaled) shear rate $\dot\gamma^{*}$ for $\varphi=0.30$ and two different values of the restitution coefficient $e$: $e=1$ and $e=0.9$. The solid and dashed lines correspond to the (perturbative) theoretical results obtained in the zeroth-order (denoted by 0th in the legend) and first-order (denoted by 1st in the legend) in $\tau_T$, respectively. Symbols refer to computer simulation results.
  }
\label{fig5}
\end{figure}

\begin{figure}[htbp]
\includegraphics[width=150mm]{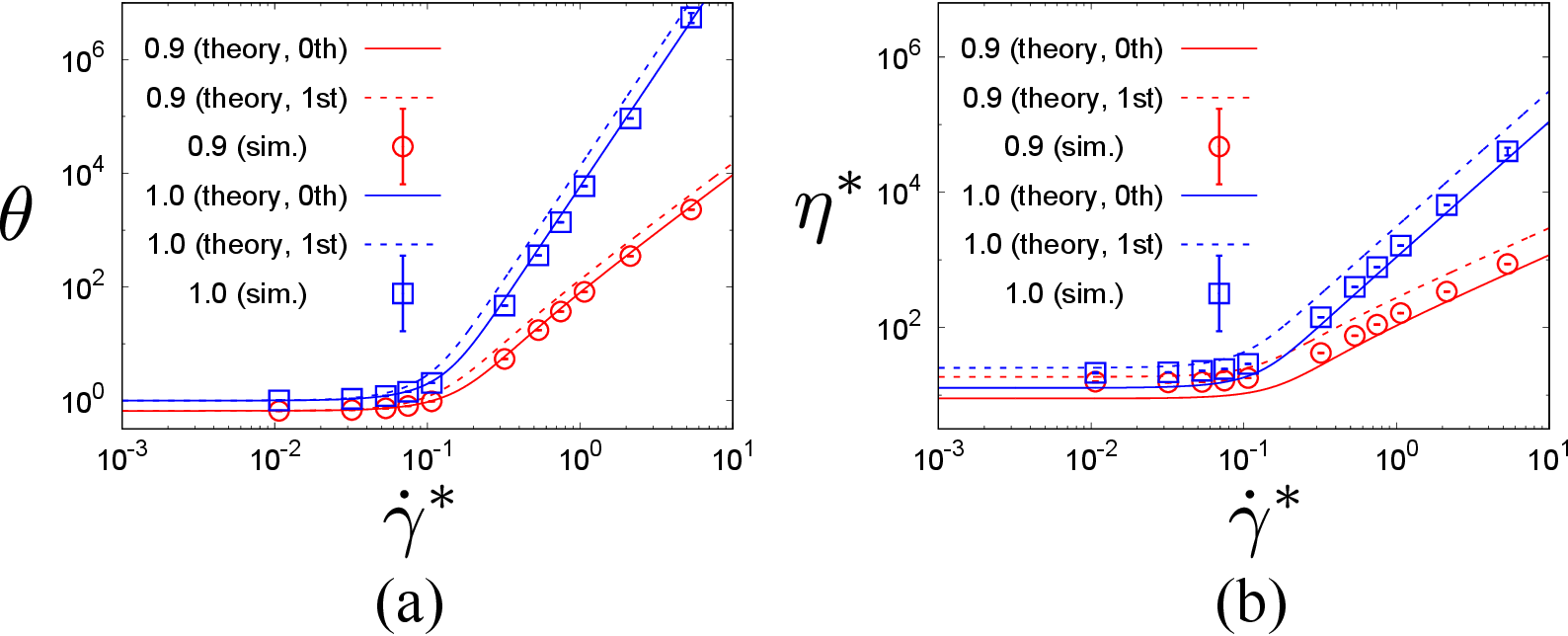}
\caption{
(Color online)
Plots of $\theta$ (panel (a)) and $\eta^*$ (panel (b)) versus the (scaled) shear rate $\dot\gamma^{*}$ for $\varphi=0.40$ and two different values of the restitution coefficient $e$: $e=1$ and $e=0.9$. The solid and dashed lines correspond to the (perturbative) theoretical results obtained in the zeroth-order (denoted by 0th in the legend) and first-order (denoted by 1st in the legend) in $\tau_T$, respectively. Symbols refer to computer simulation results.
  }
\label{fig6}
\end{figure}

\begin{figure}[htbp]
\includegraphics[width=150mm]{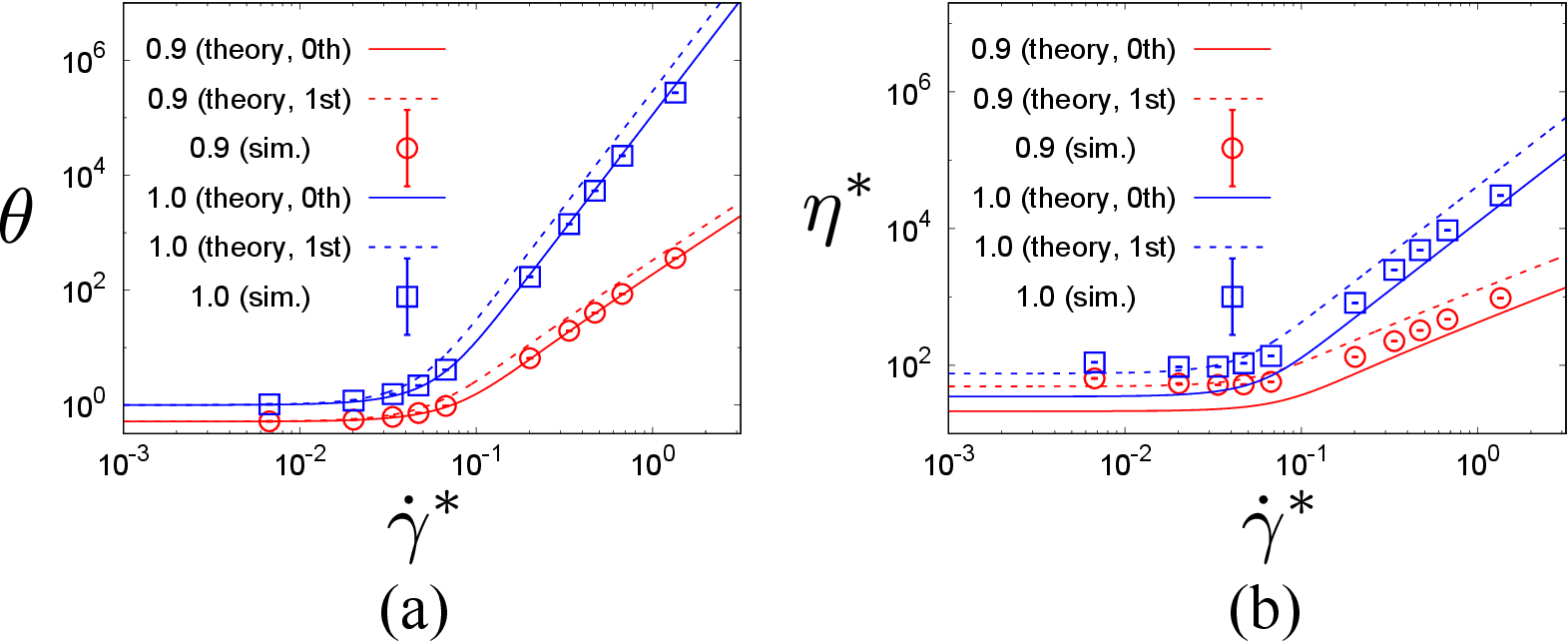}
\caption{
(Color online)
Plots of $\theta$ (panel (a)) and $\eta^*$ (panel (b)) versus the (scaled) shear rate $\dot\gamma^{*}$ for $\varphi=0.50$ and two different values of the restitution coefficient $e$: $e=1$ and $e=0.9$. The solid and dashed lines correspond to the (perturbative) theoretical results obtained in the zeroth-order (denoted by 0th in the legend) and first-order (denoted by 1st in the legend) in $\tau_T$, respectively. Symbols refer to computer simulation results.
  }
\label{fig7}
\end{figure}

Figures \ref{fig1}-\ref{fig7} show the shear-rate dependence of the (scaled) kinetic temperature $\theta$
and the (dimensionless) nonlinear shear viscosity $\eta^*$ for $\varphi=0.01$ (Fig.\ref{fig1}), $\varphi=0.05$ (Fig.\ref{fig2}), $\varphi=0.10$ (Fig.\ref{fig3}), $\varphi=0.20$ (Fig. \ref{fig4}), $\varphi=0.30$ (Fig.\ref{fig5}), $\varphi=0.40$ (Fig. \ref{fig6}) and $\varphi=0.50$ (Fig. \ref{fig7}). According to Eq.\ \eqref{shear_viscosity}, the (scaled) viscosity $\eta^*$ is defined as
\beq
\label{visc_vic}
\eta^*\equiv \frac{\zeta_0 \eta}{n T_\text{ex}}=-\frac{\theta P_{xy}^*}{\dot\gamma^*},
\eeq
where $P_{ij}^*\equiv P_{ij}/nT$. The dashed lines in those plots correspond to the (perturbative) theoretical results obtained by retaining the first-order terms in $\tau_T$ [namely, when the (scaled) shear rate is approximated by $\dot\gamma^*=\dot\gamma_0+\dot\gamma_1 \tau_T$].
These results will be referred here to as the first-order theory. Analogously, the solid lines refer to the theoretical results by assuming $\tau_T=0$ (zeroth-order theory).
We recall that the term proportional to $\dot\gamma^*\tau_T$ is the last term appearing in the expression \eqref{P_c:main_text} for $P_{\alpha\beta}^c$.
Moreover, the symbols in Figs.\ \ref{fig1}-\ref{fig7} correspond to the simulation results. Surprisingly, we observe that in general the zeroth-order results compare better with simulations than the first-order results.
On the other hand, as expected, both theories (zeroth- and first-order theories) are practically indistinguishable for dilute suspensions (see Figs.\ \ref{fig1} and \ref{fig2}).
Regarding the comparison with simulations, it is quite apparent that the zeroth-order theoretical results for the kinetic temperature $\theta$ agree well with simulations in the complete range of densities studied. This shows the accuracy of Grad's approximation to capture the shear-rate dependence of $\theta$, even for high densities. On the other hand, although the agreement between theory and simulation for $\eta^*$ is still good for $\varphi \lesssim 0.4$, some quantitative discrepancies are observed for the highest density $\varphi=0.5$. It is interesting to note that the simulation data for viscosity in the low shear (Newtonian) regime of the high density regions ($\varphi=0.50$ and $0,40$) seem to deviate from the theoretical predictions. We believe that this deviation is originated from the crystallization which takes place at $\varphi_c=0.49$.

\begin{figure}[htbp]
\includegraphics[width=150mm]{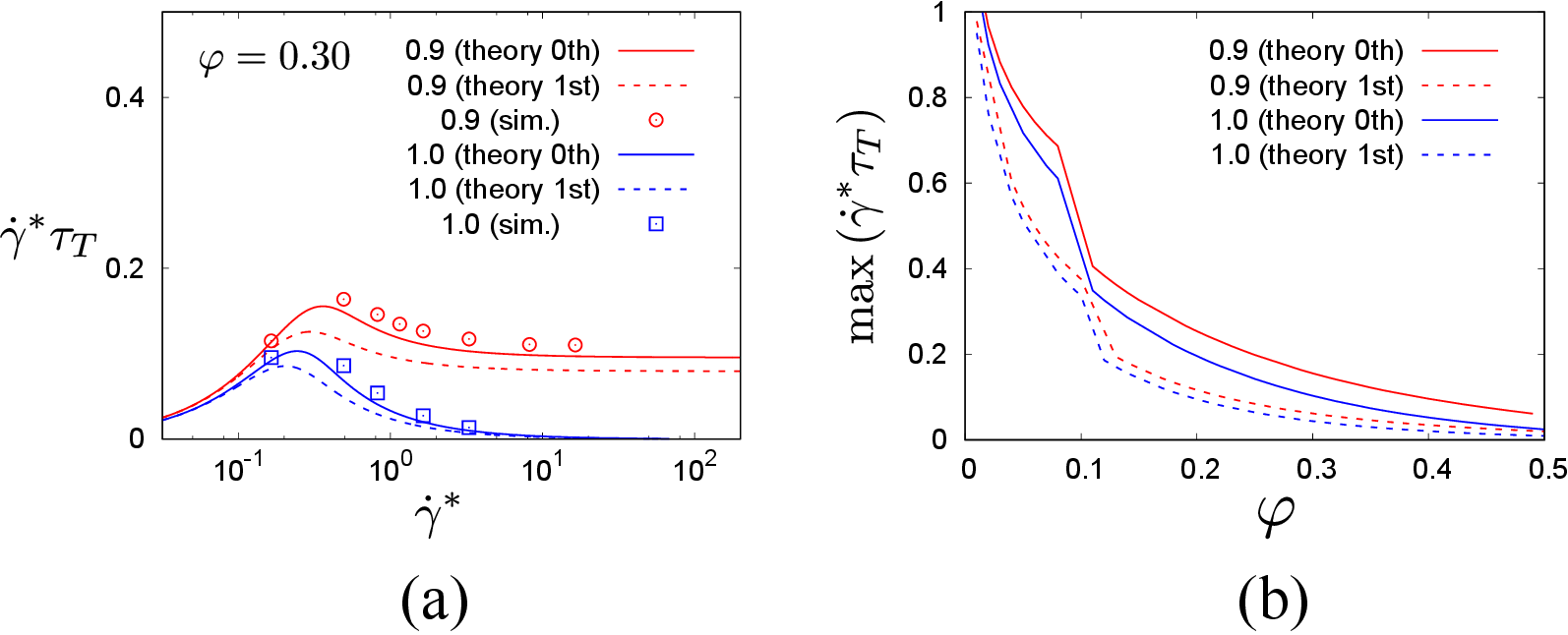}
\caption{
(Color online) Plot of $\dot\gamma^*\tau_T$ versus the (scaled) shear rate $\dot\gamma^{*}$ (panel (a)) for $\varphi=0.30$ and two different values of the restitution coefficient $e$: $e=1$ and $e=0.9$. The solid and dashed lines correspond to the (perturbative) theoretical results obtained in the zeroth-order (denoted by 0th in the legend) and first-order (denoted by 1st in the legend) in $\tau_T$, respectively. Symbols refer to computer simulation results.
Plot of the maximum value of $\dot\gamma^*\tau_T$ against the solid volume fraction (panel (b)) for $e=1$ and $e=0.9$ with the condition $\varphi\ge 0.02$.
}
\label{fig_tauT}
\end{figure}

%
As advanced in Sec. II, the evaluation of $P^c_{\alpha\beta}$  by including the complete nonlinear dependence on the shear rate yields a quite intricate expression that must be numerically integrated (see Eq.\ (3.14) of Ref.\ \cite{MGSB99}).
For this reason, a more simplified expression of $P^c_{\alpha\beta}$ has been obtained in Eq.\ \eqref{P_c:main_text} by considering only the linear contributions in the (scaled) shear rate $\dot\gamma^*$.
 On the other hand, as the panel (a) of Fig.\ \ref{fig_tauT} shows, the term $\dot\gamma^*\tau_T \propto \dot\gamma/\sqrt{T}$
becomes small in the limit of large shear rates for perfectly elastic collisions ($e=1$). This means that the contribution to the
collisional contribution to the shear stress coming from the term proportional to $\dot\gamma^*\tau_T$ in Eq.\ \eqref{P_c:main_text}
can be neglected in the case of dense gas-solid elastic suspensions.
Note that the
parameter $\dot\gamma^*\tau_T$ increases first with increasing the shear rate, reaches a maximum value and then decreases as $\dot\gamma^*$ increases. In fact, $\dot\gamma^*\tau_T$ tends asymptotically towards a constant value in the limit of large shear rates  ($\dot\gamma^*\to \infty$) for inelastic collisions [see Fig.~\ref{fig_tauT} (a) for $\varphi=0.3$].
The maximum value of $\dot\gamma^*\tau_T$ (which occurs at the (scaled) shear rate $\dot\gamma^*=\dot\gamma_\tau$) is obtained from the condition
\begin{equation}
\left(\frac{\partial (\dot\gamma^*\tau_T)}{\partial \dot\gamma^*}\right)_{\dot\gamma^*=\dot\gamma_\tau}=0.
\end{equation}
The dependence of $\text{max}(\dot\gamma^*\tau_T)$ on the solid volume fraction $\varphi$ is plotted in the panel (b) of Fig.\ \ref{fig_tauT} for $e=1$ and $e=0.9$. It is quite apparent that $\text{max}(\dot\gamma^*\tau_T)$ decreases as $\varphi$ increases. Since the collisional contribution $\mathsf{P}^c$ to the shear stress decreases with increasing the density, then one can conclude that $\mathsf{P}^c$ displays a weak dependence on the parameter $\dot\gamma^*\tau_T$ in the complete range of $\varphi$, at least for not quite high
inelasticity. This is likely the main reason for which the approximation $\tau_T=0$ in the collisional stress gives good results for $\theta$ and $\eta^*$.

Figures\ \ref{fig1}-\ref{fig7} clearly highlight that both theory and simulation predict that both $\theta$ and $\eta^*$ monotonically increase with $\dot\gamma^*$ from the Newtonian branch in the low shear regime to the Bagnolian branch for $e<1$ or the branch in which the viscosity is proportional to $\dot\gamma^2$ for $e=1$ in the high shear regime for densities $\varphi\gtrsim 0.05$.
Similar CST for dense suspensions of $e=1$ has been observed in Ref.~\cite{Kawasaki14}.
On the other hand, these monotonic tendencies disagree with the shear thinning effect observed in dense disordered suspensions in the low shear regime.
This might suggest that the shear thinning could be suppressed if one would use a mono-disperse suspension.
On the other hand, the flow curves have S-shapes for the dilute suspension $\varphi=0.01$.
More precisely, the shear thickening is continuous (CST) above the critical volume fraction $\varphi_c\approx 0.0176$, while it is discontinuous (DST) for $\varphi<\varphi_c$.
This is an interesting finding that contrasts with typical experimental observations for dense suspensions.
Notice that a similar change from a discontinuous transition to a continuous transition for the kinetic temperature has already been reported in Refs.~\cite{BGK2016,Saha17,Sangani96}.
The detailed theoretical explanation of this discontinuous-continuous transition will be presented in the next section. 
As occurs in driven granular fluids \cite{Garzo13b}, we also observe the weak influence of inelasticity on $\theta$ and $\eta^*$ for small shear rates. This is because the influence of the interstitial fluid (accounted for by the thermostat and the viscous damping term) on the dynamics of grains is more important than the effect of collisions in the low shear regime. On the other hand, the impact of inelasticity on rheology increases with increasing the shear rate.

\begin{figure}[htbp]
\includegraphics[width=150mm]{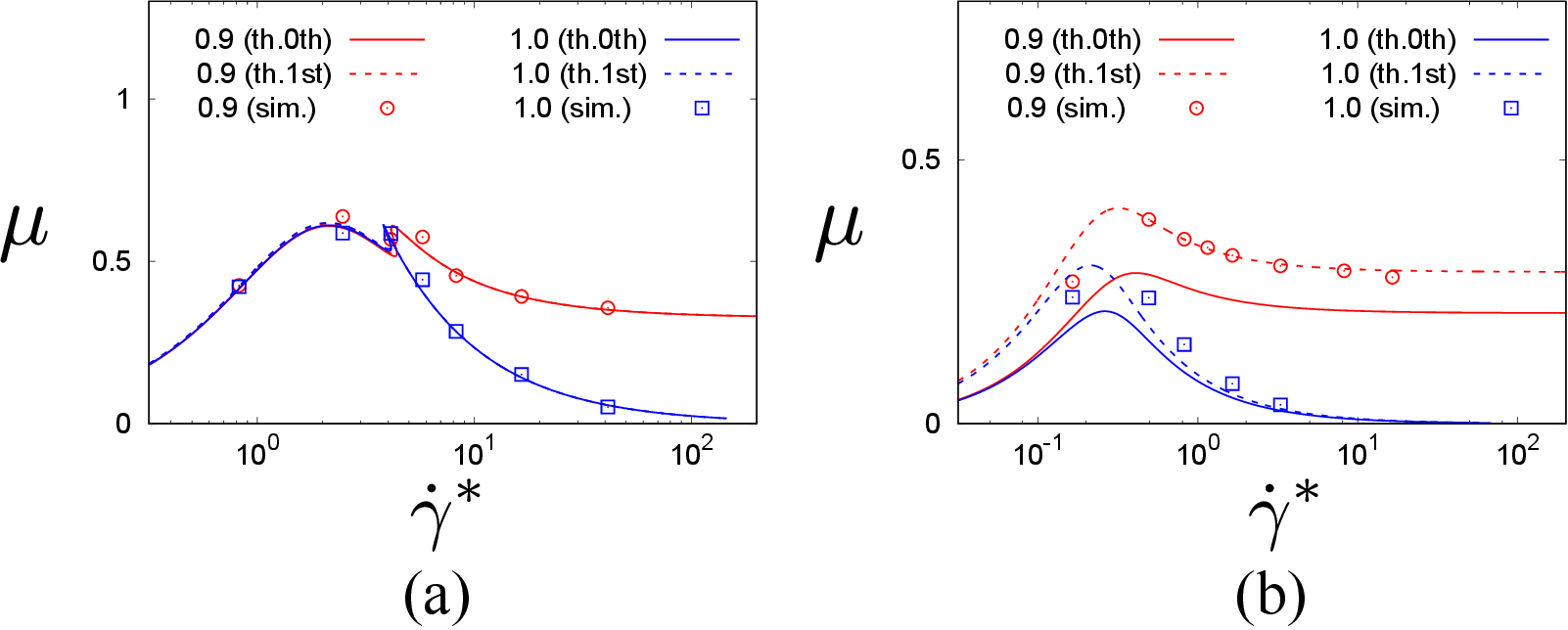}
\caption{
(Color online)
Plots of the stress ratio $\mu\equiv -P_{xy}/P$ versus the (scaled) shear rate $\dot\gamma^{*}$ for $\varphi=0.01$ (panel (a)) and $\varphi=0.30$ (panel (b)) and two different values of the restitution coefficient $e$: $e=1$ and $e=0.9$. The solid and dashed lines correspond to the (perturbative) theoretical results obtained in the zeroth-order (denoted by 0th in the legend) and first-order (denoted by 1st in the legend) in $\tau_T$, respectively. Symbols refer to computer simulation results.
}
\label{fig_mu1}
\end{figure}

Now, the results of the shear-rate dependence of the stress ratio $\mu\equiv -P_{xy}/P$ are presented in Fig.~\ref{fig_mu1}. The panel (a) of Fig.\ \ref{fig_mu1} shows the theoretical results of the dilute case ($\varphi=0.01$), where the theory gives almost perfect agreement with simulations.
The asymptotic expression of $\mu$ for large $\dot\gamma^*$ strongly depends on whether collisions are elastic or inelastic.
 In particular, while the stress ratio reaches a plateau when $e<1$, $\mu$ tends to zero in the limit $\dot\gamma^*\to \infty$ when $e=1$ as explained in Ref.~\cite{DST16}.
The results of $\mu$ for denser situations are interesting (see the panel (b) of Fig.~\ref{fig_mu1} for $\varphi=0.30$)
because the first-order theory compares better with simulations than the simple results with $\tau_T=0$.
This result contrasts with the findings of Fig.~\ref{fig5} where the zeroth-order theory provides the best performance.
This change of behavior can be understood because although the zeroth-order theory for both $P$ and $P_{xy}$ deviates
from the simulation data less than the first-order one, the opposite happens for the ratio $\mu=-P_{xy}/P$
due to a cancelation of errors. See the Appendix \ref{app:mu} for details on this point.

\begin{figure}[htbp]
 \begin{tabular}{cc}
\centering
\includegraphics[width=90mm]{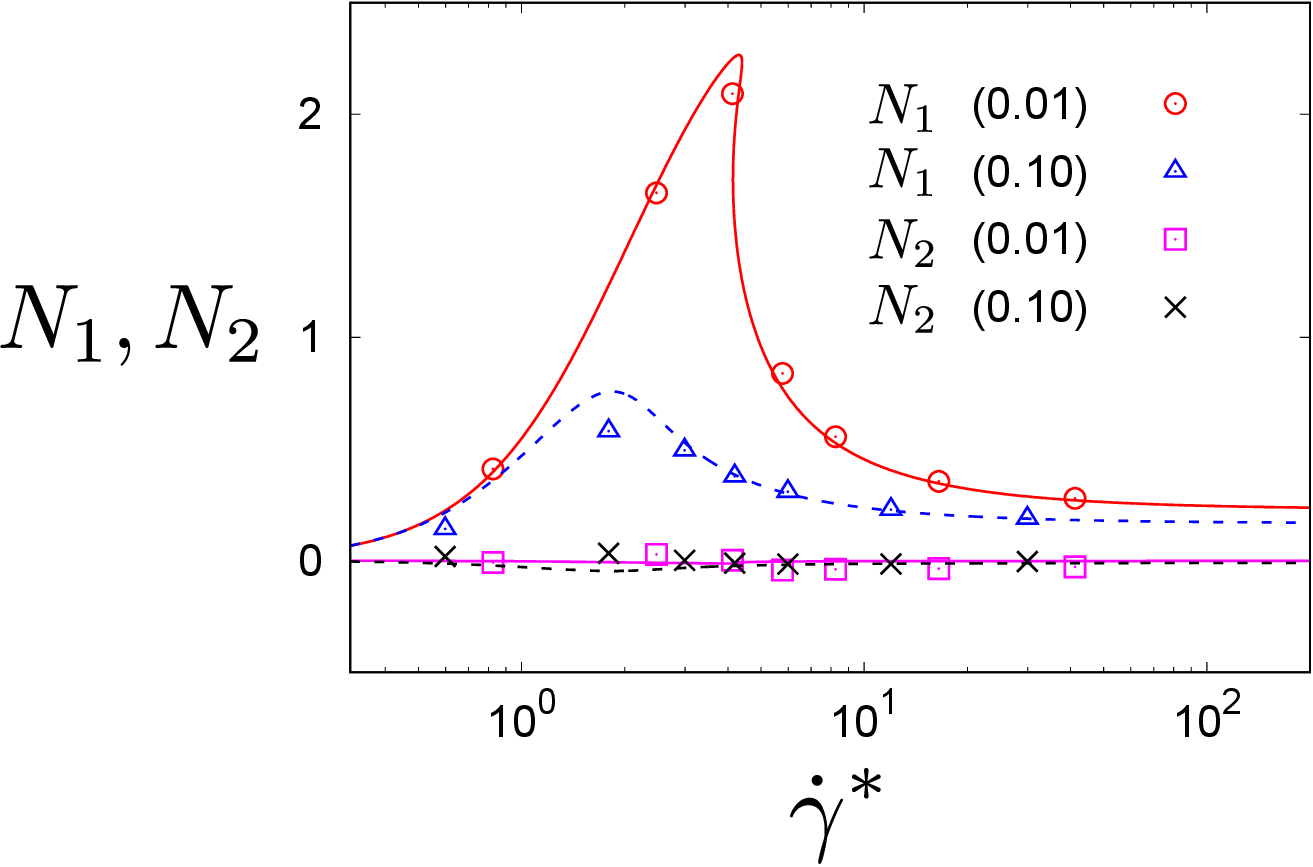}
\end{tabular}
\caption{
(Color online)
Plots of the scaled normal stress differences $N_1$ and $N_2$ versus the (scaled) shear rate $\dot\gamma^*$ from the kinetic theory and those from the simulation against $\dot\gamma^*$ for $e=0.9$ and two different values of the solid volume fraction: $\varphi=0.01$ and $0.10$. The solid and dashed lines are the theoretical results obtained by assuming $\tau_T=0$ while the symbols correspond to the simulation results.
  }
\label{N1N2_comp}
\end{figure}

 We consider now the normal stress differences $N_1$ and $N_2$. They are defined as
\beq
\label{normal.1}
N_1\equiv \frac{P_{xx}-P_{yy}}{P}, \quad N_2\equiv \frac{P_{yy}-P_{zz}}{P}.
\eeq
In terms of $\Delta \theta$ and $\delta \theta$, the expressions of $N_1$ and $N_2$ are
\beq
\label{normal.2}
N_1=\displaystyle\frac{1+\frac{2^{d-1}}{d+2}(1+e)\varphi g_0}{1+2^{d-2}(1+e)\varphi g_0}\frac{\Delta \theta}{\theta},
\eeq
\beq
\label{normal.3}
N_2 =\displaystyle\frac{1+\frac{2^{d-1}}{d+2}(1+e)\varphi g_0}{1+2^{d-2}(1+e)\varphi g_0}\frac{\delta \theta-\Delta \theta}{\theta}.
\eeq
Figure \ref{N1N2_comp} shows $N_1$ and $N_2$ versus $\dot\gamma^*$ for $e=0.9$ and two different solid volume fractions $\varphi$: $\varphi=0.01$ (dilute suspensions) and 0.1 (moderately dense suspension).
 Only the theoretical results of the zeroth-order approximation are plotted. It is seen that the theory agrees well with simulations for this range of densities.
On the other hand, the deviations between theory and simulations becomes larger for higher densities.
Moreover, it must be stressed that the normal stress differences become large when the shear thickening takes place. In particular, such a tendency is clearly observed if we focus on $N_1$ in the vicinity of the critical shear rate of the DST for dilute suspensions.

\section{Transition from discontinuous shear tickening (DST) to continuous shear tickening (CST)}
\label{DST}



%
The results discussed in Sec.\ \ref{simulation} have clearly provided evidence on the fact that the DST observed for dilute suspensions tends towards the CST as the density increases.
This transition can be analyzed as follows. For simplicity, we focus in this section on the discontinuous-continuous transition for the kinetic temperature between an ignited state and a quenched state. This transition is almost equivalent to the one found between the DST and the CST.

Because we are interested in a constant volume system, the condition for obtaining the critical point is given by 
\begin{equation}
\label{critical_cd}
\left(\frac{\partial \dot\gamma^*}{\partial \theta} \right)_{e,\varphi}=0,
\quad  {\rm and} \quad
\left(\frac{\partial^2 \dot\gamma^*}{\partial \theta^2} \right)_{e,\varphi}=0.
\end{equation}
This condition is analogous to that of the critical point of the second-order phase transition at equilibrium.

Let us determine the critical point. %
In order to get it, we consider the zeroth-order theory and so,
\beq
\label{zeroth}
\dot \gamma^{*2}=-\frac{1}{R(\varphi)^2}\frac{\mathscr{C}_0(e,\varphi,\theta)}{\mathscr{C}_2^{(0)}(e,\varphi,\theta)}.
\eeq
From Eq.~\eqref{zeroth}, the conditions \eqref{critical_cd} can be rewritten as
\begin{eqnarray}
\label{critical_cd2}
\left(\frac{\partial \mathscr{C}_0}{\partial \theta}\right)_\varphi \mathscr{C}_2^{(0)}
	-\mathscr{C}_0\left(\frac{\partial \mathscr{C}_2^{(0)}}{\partial \theta}\right)_{e,\varphi} &=&0,
\\
\left(\frac{\partial^2 \mathscr{C}_0}{\partial \theta^2}\right)_\varphi \mathscr{C}_2^{(0)}
	-\mathscr{C}_0\left(\frac{\partial^2 \mathscr{C}_2^{(0)}}{\partial \theta^2}\right)_{e,\varphi} &=&0.
\label{critical_cd2_2}	
\end{eqnarray}
For a given value of the restitution coefficient $e$, the numerical solution to Eqs.~\eqref{critical_cd2} and \eqref{critical_cd2_2} provides the critical point.
In particular, for elastic collisions ($e=1$), the critical point is given by $\varphi_{\rm c}\simeq 0.0176$, $\theta_{\rm c}\simeq38.4$, and $\dot\gamma_{\rm c}\simeq4.39$.

\begin{figure}[htbp]
\includegraphics[width=150mm]{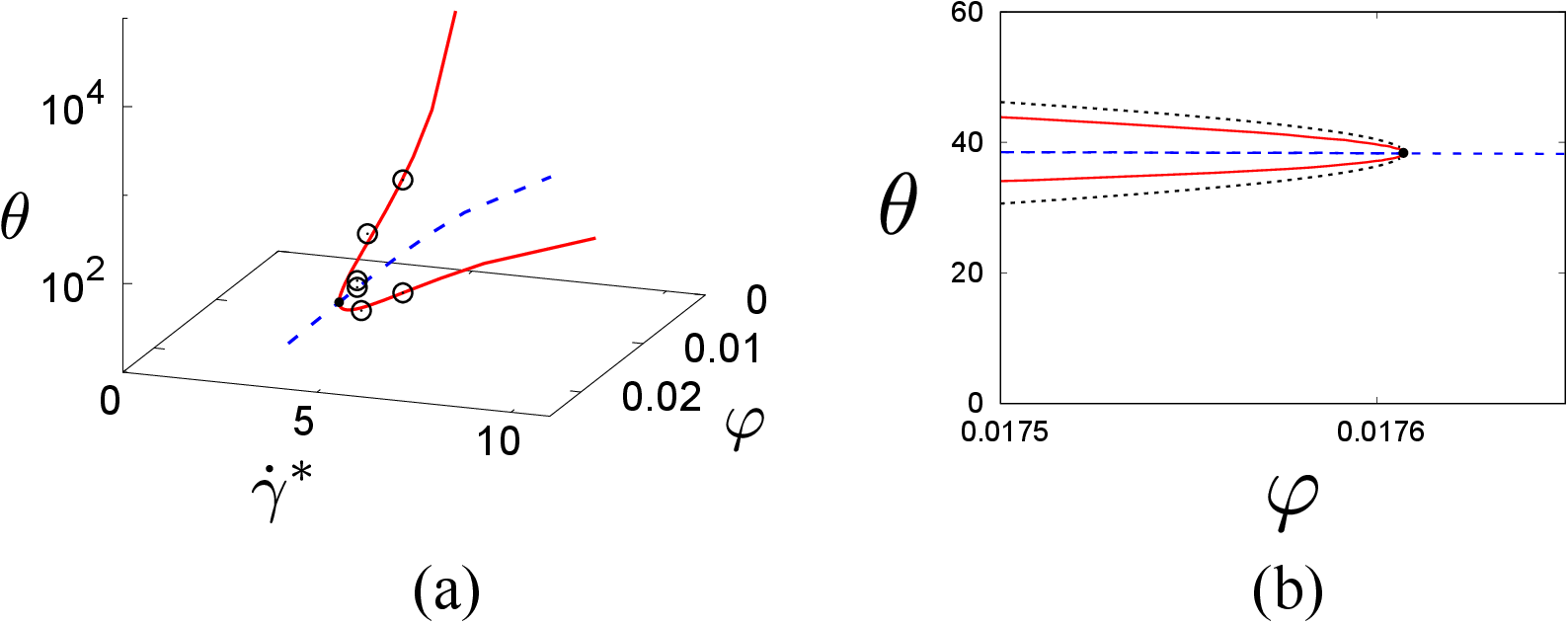}
\caption{
(Color online) Panel (a) Plots of the phase coexistence line $\partial\dot\gamma/\partial \theta=0$ (solid lines) and the spinodal line $\partial^2\dot\gamma/\partial \theta^2=0$ (dashed line).
We also plot the results of our simulation (open circles), where the temperature discontinuously increases (decreases) when we gradually increase (decrease) the shear rate.
Notice that the phase coexistence curve does not exist for $\varphi>\varphi_c$.
Panel (b) Plots of the projection of the phase coexistence line and the spinodal line onto the $(\varphi,\theta)$-plane.
}
\label{critical}
\end{figure}

As the panel (a) of Fig.\ \ref{critical} shows, Eqs.\ \eqref{critical_cd2} and \eqref{critical_cd2_2} can be seen as analogous to the phase coexistence and spinodal lines at equilibrium phase transitions, respectively, in the phase space of $(\theta, \varphi,\dot\gamma^*)$. Because of this analogy, we will employ the above terminology for the later discussion.

To confirm the validity of our analysis, we have also performed the EDLSHS simulations in the vicinity of the critical point for the case $e=1$.
We have gradually changed the shear rate from $\dot\gamma^*_0=0.400 (0.826)$ to sequentially increasing (decreasing) values as $\dot\gamma^*=\dot\gamma^*_0, a\dot\gamma^*_0, a^2\dot\gamma^*_0, \cdots, a^{63}\dot\gamma^*_0=0.826 (0.400)$ with the rate $a=10^{0.005}\simeq1.0116$.
We have verified that the coexistence of an ignited state and a quenched state in our simulation exists on the phase coexistence line as shown in the panel (a) of Fig.~\ref{critical}.
The intersection of the two lines correspond to the critical point.
Notice that the spinodal line is located outside the phase coexistence line in our case, which is different from equilibrium situations.
This difference might be a universal feature of non-equilibrium bifurcations because models of traffic flows have similar structures~\cite{Komatsu95,Hayakawa98}.

Near the critical point, the equation of the coexistence curve between $\theta-\theta_c$ and $\varphi_c-\varphi$ for $\varphi<\varphi_c$ is determined as
\begin{equation}
\label{theory}
\theta-\theta_c=\pm C \sqrt{\varphi_c-\varphi} ,
\end{equation}
where $C=\left\{6(\partial^2 \dot\gamma^*/\partial\theta\partial\varphi)/(\partial^3\dot\gamma^* /
\partial \theta^3)\right\}_{\varphi_{\rm c},\theta_{\rm c}}^{1/2}\simeq 750$ for $e=1$.
The theoretical curve in Eq.~\eqref{theory} is drawn as the solid (red) line in  the panel (b) of Fig.~\ref{critical}.
This analytical prediction captures qualitatively well the numerical result obtained from Eqs.~\eqref{critical_cd2}
and \eqref{critical_cd2_2} (the doted line in Fig.~\ref{critical}).

%


\section{Discussion and conclusion}
\label{discussion}

The Enskog kinetic equation for inelastic hard spheres has been considered in this paper as the starting point to study the rheology of gas-solid suspensions under simple shear flow.
The effect of the interstitial fluid on the dynamics of solid particles has been modeled through an external force composed by a viscous drag force plus a stochastic Langevin-like term.
While the first term models the friction of grains on the gas phase, the latter accounts for thermal fluctuations.
Two independent but complementary  routes have been employed to determine the non-Newtonian transport properties.
First, the Enskog equation has been approximately solved by means of Grad's moment method. 
 Given that the heat flux vanishes in the simple shear flow state, only the kinetic pressure tensor has been retained in the trial distribution function.
  Then, the analytical results for the kinetic temperature, the viscosity, the stress ratio, and the normal stress differences have been compared against computer simulations based on the event-driven Langevin simulation method.
The main goal of the paper has been to determine how the flow curve (stress-strain rate relation) depends on the density (or volume fraction) of the confined gases.

One of the limitations of the theory is that the collisional moment $\overline{\Lambda}_{\alpha\beta}^E$ [defined by Eq.\ \eqref{over_Lambda}] has been evaluated by neglecting nonlinear terms in the kinetic pressure tensor $\Pi_{\alpha\beta}^k$. 
For dilute gases ($\varphi\to 0$), this simplification leads to the absence of normal stress differences in the shear flow plane ($P_{xx}^k= P_{yy}^k$). However, although this equality differs from the results found in computer simulations \cite{Tsao95,Chamorro15}, the difference $P_{xx}^k-P_{yy}^k$ observed in simulations is in general very small.
 As a consequence, the importance of this approximation seems to be not relevant for the calculations carried out in the present paper. Another simplification of our theory is that one of the contributions to the collisional stress $P_{xy}^c$ has been determined by neglecting nonlinear terms in the shear rate [see the third term on the right hand side of Eq.\ \eqref{P_c:main_text}]. 
 On the other hand, the comparison with simulations has shown that the reliability of the theory is clearly improved when this term is neglected (zeroth-order theory).

The theoretical results derived in this paper from Grad's method indicate that in general the Enskog theory describes well the rheology of sheared suspensions.
In particular, the agreement found between theory and simulations for the shear viscosity clearly shows that the shear thickening effect is well captured by the Enskog kinetic equation.
Moreover, in contrast to typical experimental observations for dense suspensions, both theory and simulations have confirmed that there is a transition from the DST in dilute suspensions to the CST for dense suspensions at relatively low density.
This finding is consistent with the results reported in previous works ~\cite{DST16,Tsao95,BGK2016,Sangani96,Santos04,Saha17} where only the transition between the quenched state and the ignited state for the kinetic temperature was analyzed.

As advanced before, in spite of the fact that our theoretical results are based in some approximations, it must be stressed that the theoretical predictions for the shear-rate dependence of the shear viscosity compare well with simulations for moderately dense suspensions (for instance, densities $\varphi$ smaller than or equal to 0.3).
This is the expected result since several previous works \cite{LBD02,DHGD02,MGAL06,MDCPH11,MGH14} have confirmed the reliability of the Enskog equation in this range of densities.
The disagreement between theory and simulation for denser cases could be in part originated by the incomplete treatment of the collisional stress ${\sf P}^c$ where our expression is the same as the one obtained by Garz\'{o} and Dufty~\cite{Garzo99} from the first-order Chapman-Enskog solution. Given that the latter theory is not applicable in the high shear-rate regime, it is obvious that the present results could be refined by considering higher-order terms in the shear rate in the expression of the collisional stress.
This point is one of the important tasks for the near future.

Typical DSTs observed in experiments and simulations for dense suspensions ($\varphi>0.5$) should be the result of mutual friction between grains.
Although the Enskog kinetic equation is not applicable to such dense suspensions, an extension of Grad's moment method to dense systems might be applicable for the explanation of the DST of frictional grains~\cite{Suzuki17},
 which might be better than the previous theory of dense granular liquids~\cite{Suzuki15}.
This study will be reported elsewhere~\cite{Saitoh17} (see also Ref.~\cite{Saitoh16}).

The Langevin equation \eqref{Langevin_eq} employed in our study assumes that the gravity force is perfectly balanced with the drag force immersed by the air flow. This assumption is only true if the homogeneous state is stable. On the other hand, the simple shear flow state becomes unstable above the critical shear rate.
If the homogeneous state is unstable, one would need to consider the time evolution of local structure as well as the consideration of the inhomogeneous drag.

The fact that the restitution coefficient $e$ is assumed to be constant has allowed to get quite explicit results.
However, the above hypothesis disagrees with experimental observations \cite{BHL84} or with mechanics of particle collisions \cite{RPBS99} and hence, the coefficient $e$ depends on the impact velocity. The simplest model that takes into account dissipative material deformation is the model of viscoelastic particles \cite{BP00,BP03,DBPB13}. 
On the other hand, in spite of the mathematical difficulties involved in this viscoelastic model, some progresses have been made in the past few years \cite{BP00,BP03,DBPB13} in the limit of small inelasticity for dilute granular gases. The extension of the present results for a velocity dependent restitution coefficient is beyond the scope of this paper. In addition, 
since the transition between DST to CST for elastic suspensions is qualitatively similar to that of inelastic suspensions (except in the high shear asymptotic region), we think that the impact of the velocity dependence of $e$ on the above transition will be not relevant for such a problem.

As shown in the Appendix G, since the theoretical predictions deviate from simulation results for strong inelasticity, the reliability of our theory is essentially limited to moderate inelasticities. 
Thus, as a future task, we plan to improve our theoretical treatment for highly inelastic cases. 
Finally, it is important to note that the monodisperse system analyzed here is crystallized, at least, in the region of low shear rates for densities $\varphi>0.49$. Therefore, one should study a sheared polydisperse system to prevent it from crystallization. This is also an interesting problem to be carried out in the future.

\acknowledgements

We thank Satoshi Hayakawa, Koshiro Suzuki, Takeshi Kawasaki, Michio Otsuki, and Kuniyasu Saitoh for their useful comments.
The research of HH and ST has been partially supported by the Grant-in-Aid of MEXT for Scientific Research (Grant No. 16H04025) and the YITP activity (YITP-W-16-14).
The research of VG has been supported by the Spanish Government through Grant No. FIS2016-76359-P, partially financed by FEDER funds and by the Junta de Extremadura (Spain) through Grant No. GR15104.


\appendix



\section{Brief note on Enskog's approximation}
\label{Enskog_base}

The basis of Enskog's approximation is briefly summarized in this Appendix. Notice that the main part of this Appendix has been presented in Ref.~\cite{MGSB99}. The collisional integral $J_E(\bm{r},\bm{v}|f)$ accounting for the effect of collisions on the rate of change of the one-particle distribution function $f$ is assumed to be the inelastic hard-core collision operator. It is given by
\begin{equation}
\label{J(V|f)_app}
J_E(\bm{r},\bm{v}_1|f^{(2)})
=\sigma^{d-1}\int d\bm{v}_2\int d\hat{\bm{\sigma}}\Theta(\bm{v}_{12}\cdot\hat{\bm{\sigma}})
(\bm{v}_{12}\cdot\hat{\bm{\sigma}})
\left[
\frac{f^{(2)}(\bm{r},\bm{r}-\bm{\sigma},\bm{v}_1^{''},\bm{v}_2^{''};t)}{e^2}-f^{(2)}(\bm{r},\bm{r}+\bm{\sigma},
\bm{v}_1,\bm{v}_2;t)
\right],
\end{equation}
where $f^{(2)}\equiv f^{(2)}(\bm{r}_1,\bm{r}_2,\bm{v}_1,\bm{v}_2;t)$ is the two-body distribution function at $(\bm{r}_i,\bm{v}_i)$ with $i=1,2$. The relationship between the pre- and post-collisional velocities in Eq.\ \eqref{J(V|f)_app} is given Eq.\ \eqref{collision_rule}.

The most important flux in the simple shear flow problem is the pressure tensor $\sf{P}(\bm{r},t)$. Its kinetic $\mathsf{P}^k$ and collisional $\sf{P}^c$ contributions are, respectively, given by (see the Appendix \ref{details_collision_transfer} for the derivation):
\begin{eqnarray}
P^k_{\alpha\beta}(\bm{r},t)&=&m \int d\bm{v} V_\alpha V_\beta f(\bm{V},t) ,
\label{pressure_tensor:kinetic}
\\
P^c_{\alpha\beta}(\bm{r},t)&=&
\frac{1+e}{4}m\sigma^d\int d\bm{v}_1\int d\bm{v}_2 \int d\hat{\bm{\sigma}}
\Theta(\hat{\bm{\sigma}}\cdot\bm{v}_{12})(\hat{\bm{\sigma}}\cdot\bm{v}_{12})^2
\hat{\sigma}_\alpha \hat{\sigma}_\beta \int_0^1\; dx\; f^{(2)}[\bm{r}-x\bm{\sigma},\bm{r}+(1-x)\bm{\sigma},\bm{v}_1,\bm{v}_2;t].
\nonumber\\
\label{pressure:collison}
\end{eqnarray}

In order to achieve a closed kinetic equation for the distribution function $f$, one assumes the molecular chaos hypothesis and hence, the two-body distribution function $f^{(2)}$ factorizes in the product of the one-particle distribution functions $f$ as
\begin{equation}
f^{(2)}(\bm{r}_1,\bm{r}_2,\bm{v}_1,\bm{v}_2;t)=\chi(\bm{r}_1,\bm{r}_2|n(t))f(\bm{r}_1,\bm{v}_1,t)f(\bm{r}_2,\bm{v}_2,t) ,
\label{decoupling}
\end{equation}
where the front factor $\chi(\bm{r}_1,\bm{r}_2|n(t))$ is reduced to
the radial distribution function $g(|\bm{r}_1-\bm{r}_2|,n)$ for the simple shear flow state. Because we are only interested in systems consisting of hard spheres, $\chi(\bm{r}_1,\bm{r}_2|n(t))$ is further simplified to $\chi(\bm{r}_1,\bm{r}_2|n(t))\approx g_0(|\bm{r}|=\sigma,\varphi)$, where the radial distribution at contact $g_0(|\bm{r}|=\sigma,\varphi)$ can be expressed as in Eq.\ \eqref{radial_fn} for $d=3$ and $\varphi<0.49$~\cite{CS}. Once the Enskog approximation is adopted, the Enskog collision operator $J_E(\bm{r},\bm{V}|f^{(2)})$ can be rewritten as in Eq.~\eqref{J(V|f)} when one considers the Lagrangian frame defined by $\bm{V}=\bm{v}-\dot\gamma y \bm{e}_x$.

Moreover, to get Eq.\ \eqref{pressure:collisonal} for $\mathsf{P}^c$, one takes first the Enskog approximation \eqref{decoupling} for $f^{(2)}$ and then expands $f(\bm{r}+y\bm{\sigma})$ in spatial gradients as
\begin{eqnarray}
\label{int_f_2}
\int_0^1dx f^{(2)}(\bm{r}-x\bm{\sigma},\bm{r}+(1-x)\bm{\sigma},\bm{v}_1,\bm{v}_2;t)
&\approx&
g_0
\left[f(\bm{r},\bm{v}_1)f(\bm{r},\bm{v}_2)-\frac{1}{2}f(\bm{r}.\bm{v}_2)\bm{\sigma}\cdot\nabla f(\bm{r},\bm{v}_1)+ \frac{1}{2}f(\bm{r},\bm{v}_1)\bm{\sigma}\cdot\nabla f(\bm{r},\bm{v}_2)\right]
\nonumber\\
&\approx&
g_0\left[f(\bm{r}-\frac{\bm{\sigma}}{2},\bm{v}_1;t)f(\bm{r}+\frac{\bm{\sigma}}{2},\bm{v}_2;t) \right].
\end{eqnarray}
The expression ~\eqref{pressure:collisonal} for $\mathsf{P}^c$ can be easily obtained by substituting Eq.\ \eqref{int_f_2} into Eq.\ \eqref{pressure:collison} and referring the velocities of the particles to the local Lagrangian frame where $f$ is spatially uniform. This means that
\beq
f(\bm{r}-\frac{\bm{\sigma}}{2},\bm{v}_1;t)=f\left(\bm{V}_1+\frac{1}{2}\dot\gamma \sigma \widehat{\sigma}_y \bm{e}_x;t\right), \quad
f(\bm{r}+\frac{\bm{\sigma}}{2},\bm{v}_1;t)=f\left(\bm{V}_1-\frac{1}{2}\dot\gamma \sigma \widehat{\sigma}_y \bm{e}_x;t\right).
\eeq

\section{Some details of the collisional transfer contributions to the fluxes}\label{details_collision_transfer}

Some technical details on the derivation of the collisional transfer contributions to the fluxes are provided in this Appendix.
Notice that the description in this Appendix is applicable for all the systems of hard core collisions.
In other words, we do not use any specific property either the Enskog approximation \eqref{decoupling} or Grad's distribution \eqref{Grad}.

Let us consider the following collisional moment of the Enskog operator
\beq
I_\psi=\int d\bm{v}\; \psi(\bm{v})\; J_E(\bm{r},\bm{v}|f,f),
\eeq
where $\psi(\bm{v})$ is an arbitrary function of $\bm{v}$. The moment $I_\psi$ can be written in the equivalent form \cite{Garzo99}
\beqa
\label{Garzo1.188}
I_\psi
&=&
\sigma^{d-1}\int d\bm{v}_1\int d\bm{v}_2\int d\hat{\bm{\sigma}}
\Theta(\hat{\bm{\sigma}}\cdot\bm{v}_{12})(\hat{\bm{\sigma}}\cdot\bm{v}_{12})
[\psi(\bm{v}_1')-\psi(\bm{v}_1)]f^{(2)}(\bm{r},\bm{v}_1,\bm{r}+\bm{\sigma},\bm{v}_2;t)
\nonumber\\
&=&
\sigma^{d-1}\int d\bm{v}_1\int d\bm{v}_2\int d\hat{\bm{\sigma}}
\Theta(\hat{\bm{\sigma}}\cdot\bm{v}_{12})(\hat{\bm{\sigma}}\cdot\bm{v}_{12})
[\psi(\bm{v}_2')-\psi(\bm{v}_2)]f^{(2)}(\bm{r},\bm{v}_2,\bm{r}-\bm{\sigma},\bm{v}_1;t),
\end{eqnarray}
where
\beq
\bm{v}_1'=\bm{v}_1-\frac{1}{2}(1+e)(\widehat{\bm{\sigma}}\cdot\bm{v}_{12})\widehat{\bm{\sigma}}, \quad
\bm{v}_2'=\bm{v}_2+\frac{1}{2}(1+e)(\widehat{\bm{\sigma}}\cdot\bm{v}_{12})\widehat{\bm{\sigma}}.
\eeq
Moreover, the last expression in Eq.\ \eqref{Garzo1.188} has been obtained by interchanging $\bm{v}_1$ and $\bm{v}_2$ and changing $\hat{\bm{\sigma}}\to -\hat{\bm{\sigma}}$. Using the identities of Eq. \eqref{Garzo1.188}, the collisonal moment $I_\psi$ can be rewritten as
\begin{eqnarray}
I_\psi&=&
\frac{\sigma^{d-1}}{2}\int d\bm{v}_1\int d\bm{v}_2\int d\hat{\bm{\sigma}}
\Theta(\hat{\bm{\sigma}}\cdot\bm{v}_{12})(\hat{\bm{\sigma}}\cdot\bm{v}_{12})
\left\{ [\psi(\bm{v}_1')-\psi(\bm{v}_1)]f^{(2)}(\bm{r},\bm{v}_1,\bm{r}+\bm{\sigma},\bm{v}_2;t)\right.
\nonumber\\
& & \left.
+[\psi(\bm{v}_2')-\psi(\bm{v}_2)]f^{(2)}(\bm{r},\bm{v}_2,\bm{r}-\bm{\sigma},\bm{v}_1;t)\right\}
\nonumber\\
&=&
\frac{\sigma^{d-1}}{2}\int d\bm{v}_1\int d\bm{v}_2\int d\hat{\bm{\sigma}}
\Theta(\hat{\bm{\sigma}}\cdot\bm{v}_{12})
(\hat{\bm{\sigma}}\cdot\bm{v}_{12})
\{[\psi(\bm{v}_1')+\psi(\bm{v}_2')-\psi(\bm{v}_1)-\psi(\bm{v}_2)]
f^{(2)}(\bm{r},\bm{v}_1,\bm{r}+\bm{\sigma},\bm{v}_2;t)
\nonumber\\
&&
\quad +[\psi(\bm{v}_1')-\psi(\bm{v}_1)]
[
f^{(2)}(\bm{r},\bm{v}_1,\bm{r}+\bm{\sigma},\bm{v}_2;t)-
f^{(2)}(\bm{r}-\bm{\sigma},\bm{v}_1,\bm{r},\bm{v}_2;t)
]
\}.
\label{Garzo1.193}
\end{eqnarray}
Upon deriving the last identity use has been of the relation
\beq
f^{(2)}(\bm{r},\bm{v}_2,\bm{r}-\bm{\sigma},\bm{v}_1;t)
=f^{(2)}(\bm{r}-\bm{\sigma},\bm{v}_1,\bm{r},\bm{v}_2;t).
\eeq
The first term in the integrand of Eq.~\eqref{Garzo1.193} on the right hand side represents a collisional effect due to a change in velocities. This effect is also present in the dilute regime.
The second term on the right hand side in the integrand of Eq.~\eqref{Garzo1.193} expresses a pure collisional contribution.
Now, we use the following identity for an arbitrary function $F(\bm{r},\bm{r}+\bm{\sigma})$:
\begin{equation}
\label{identity}
F(\bm{r},\bm{r}+\bm{\sigma})-F(\bm{r}-\bm{\sigma},\bm{r})
=-\int_0^1\; dx \frac{\partial}{\partial x}F[\bm{r}-x \bm{\sigma},\bm{r}+(1-x)\bm{\sigma}]
=\bm{\sigma}\cdot\frac{\partial}{\partial \bm{r}}\int_0^1\; dx F[\bm{r}-x \bm{\sigma},\bm{r}+(1-x)\bm{\sigma}].
\end{equation}
From the identity \eqref{identity}, Eq.\ \eqref{Garzo1.193} can be rewritten as
\begin{eqnarray}\label{Garzo1.195}
I_\psi&=&
\frac{\sigma^{d-1}}{2}\int d\bm{v}_1\int d\bm{v}_2\int d\hat{\bm{\sigma}}
\Theta(\hat{\bm{\sigma}}\cdot\bm{v}_{12})
(\hat{\bm{\sigma}}\cdot\bm{v}_{12})
\{
 [\psi(\bm{v}_1')+\psi(\bm{v}_2')-\psi(\bm{v}_1)-\psi(\bm{v}_2)]
f^{(2)}(\bm{r},\bm{v}_1,\bm{r}+\bm{\sigma},\bm{v}_2;t)
\nonumber\\
&&
\quad
+ \nabla \cdot [\psi(\bm{v}_1')-\psi(\bm{v}_1)]  \bm{\sigma}
\int_0^1\; dx f^{(2)}[\bm{r}-x \bm{\sigma},\bm{v}_1,\bm{r}+(1-x)\bm{\sigma},\bm{v}_2;t].
\end{eqnarray}

It is straightforward to show
\begin{equation}\label{I_1=0}
I_1=\int d\bm{v} J_E(\bm{V}|f,f)=0.
\end{equation}
In the case $\psi(\bm{v})=m\bm{v}$, the first term on the right hand side of Eq.~\eqref{Garzo1.195} vanishes since $\bm{v}_1'+\bm{v}_2'=\bm{v}_1+\bm{v}_2$. Therefore, the second term on the right hand side of Eq.~\eqref{Garzo1.195} yields
\begin{equation}
\label{Garzo1.196}
I_{m\bm{v}}=-\frac{1+e}{4} m\sigma^d \nabla\cdot
\int d\bm{v}_1\int d\bm{v}_2 \int d\hat{\bm{\sigma}}
\Theta(\hat{\bm{\sigma}}\cdot\bm{v}_{12})
(\hat{\bm{\sigma}}\cdot\bm{v}_{12})^2
\hat{\bm{\sigma}}\hat{\bm{\sigma}}
\int_0^1dx f^{(2)}[\bm{r}-x\bm{\sigma},\bm{v}_1,\bm{r}+(1-x)\bm{\sigma},\bm{v}_2;t] .
\end{equation}
This equation can be rewritten as
\begin{equation}
\label{Garzo1.198}
I_{m\bm{v}}=-\nabla\cdot {\sf P}^c,
\end{equation}
where ${\sf P}^c$ is given by Eq.~\eqref{pressure:collisonal}.

Now we consider the kinetic energy $\psi(\bm{v})=m v^2/2$. In this case, the first term on the right hand side of Eq.~\eqref{Garzo1.195} is different from zero since energy is not conserved in collisions. Thus, one obtains
\begin{eqnarray}
I_{mv^2/2}&=&
\frac{m\sigma^{d-1}}{8}(1-e^2)
\int d\bm{v}_1\int d\bm{v}_2\int d\hat{\bm{\sigma}}
\Theta(\hat{\bm{\sigma}}\cdot\bm{v}_{12})
(\hat{\bm{\sigma}}\cdot\bm{v}_{12})^3
f^{(2)}(\bm{r},\bm{v}_1,\bm{r}+\bm{\sigma},\bm{v}_2;t)
\nonumber\\
&&
-\nabla\cdot \frac{m\sigma^d}{4}(1+e)
 \int d\bm{v}_1\int d\bm{v}_2\int d\hat{\bm{\sigma}}
\Theta(\hat{\bm{\sigma}}\cdot\bm{v}_{12})
(\hat{\bm{\sigma}}\cdot\bm{v}_{12})^2
\hat{\bm{\sigma}}
\left[
\frac{1-e}{4}
(\hat{\bm{\sigma}}\cdot\bm{v}_{12})+(\hat{\bm{\sigma}}\cdot\bm{V}_G)+
(\hat{\bm{\sigma}}\cdot\bm{u})
\right]
\nonumber\\
&& \times
\int_0^1 dx f^{(2)}(\bm{r}-x \bm{\sigma},\bm{v}_1,\bm{r}+(1-x)\bm{\sigma},\bm{v}_2;t) ,
\label{Garzo_1.199}
\end{eqnarray}
where $\bm{V}_G\equiv (\bm{V}_1+\bm{V}_2)/2$ is the velocity of center of mass. In addition, we have employed the identities
\beq
\bm{v}_1'^2-\bm{v}_1^2=
-(1+e)(\bm{v}_{12}\cdot\hat{\bm{\sigma}})\left[(\bm{u}+\bm{V}_G)\cdot\hat{\bm{\sigma}}\right]
-(1-e^2)(\bm{v}_{12}\cdot\hat{\bm{\sigma}})^2,
\eeq
\beq
v_1'^{2}+v_2'^2-v_1^2-v_2^2=-\frac{1-e^2}{4}(\bm{v}_{12}\cdot\hat{\bm{\sigma}})^2.
\eeq
Equation \eqref{Garzo_1.199} can be rewritten as
\begin{equation}
\label{Garzo1.202}
I_{mv^2/2}=-\frac{d}{2}nT \xi-\nabla \cdot (\bm{u}: {\sf P}^c+\bm{q}^c)
-\nabla\cdot\bm{S},
\end{equation}
where
\beq
\label{xi}
\xi=(1-e^2)\frac{m \sigma^{d-1}}{4dnT}\int d\bm{v}_1\int d\bm{v}_2\int d\hat{\bm{\sigma}}\Theta(\hat{\bm{\sigma}}\cdot\bm{v}_{12})(\hat{\bm{\sigma}}\cdot\bm{v}_{12})^3 f^{(2)}(\bm{r},\bm{v}_1,\bm{r}+\bm{\sigma},\bm{v}_2;t)
\eeq
is the cooling rate,
\beq
\label{q^c}
\bm{q}^c=\frac{1+e}{4}m\sigma^d\int d\bm{v}_1 \int d\bm{v}_2\int d\hat{\bm{\sigma}}\Theta(\hat{\bm{\sigma}}\cdot\bm{v}_{12})(\hat{\bm{\sigma}}\cdot\bm{v}_{12})^2(\hat{\bm{\sigma}}
\cdot\bm{V}_G)\int_0^1 dx f^{(2)}[\bm{r}-x \bm{\sigma},\bm{v}_1,\bm{r}+(1-x)\bm{\sigma},\bm{v}_2;t]
\eeq
is the collisional contribution to the heat flux and
\beqa
\bm{S}&=& (1-e^2)\frac{m \sigma^{d-1}}{16}
\int d\bm{v}_1\int d\bm{v}_2\int d\hat{\bm{\sigma}}\Theta(\hat{\bm{\sigma}}\cdot\bm{v}_{12})(\hat{\bm{\sigma}}\cdot\bm{v}_{12})^3
 \bm{\sigma} \int_0^1 dx f^{(2)}[\bm{r}-x \bm{\sigma},\bm{v}_1,\bm{r}+(1-x)\bm{\sigma},\bm{v}_2;t]
\nonumber\\
&=&
-(1-e^2)\frac{m \sigma^{d-1}}{16}
\int d\bm{v}_1\int d\bm{v}_2\int d\hat{\bm{\sigma}}\Theta(\hat{\bm{\sigma}}\cdot\bm{v}_{12})(\hat{\bm{\sigma}}\cdot\bm{v}_{12})^3
 \bm{\sigma} \int_0^1 dx f^{(2)}[\bm{r}-x \bm{\sigma},\bm{v}_1,\bm{r}+(1-x)\bm{\sigma},\bm{v}_2;t]
\nonumber\\
&=& -\bm{S},
\eeqa
where in the last step we have exchanged $1\leftrightarrow 2$ and have made the change of variable $\bm{\sigma} \to -\bm{\sigma}$. Since $\bm{S}=-\bm{S}$, then the vector $\bm{S}=0$ if all the grains are identical. With this result, Eq.\ \eqref{Garzo1.202} reduces to
\beq
\label{energybalance}
I_{mv^2/2}=-\frac{d}{2}nT \xi-\nabla \cdot (\bm{u}: {\sf P}^c+\bm{q}^c).
\eeq
Then, the trace of the collisional moment $\Lambda_{\alpha\beta}^E$ defined in Eq.\ \eqref{vic4} can be rewritten as
\begin{equation}
\Lambda_{\alpha\alpha}^E=\bm{u}\cdot I_{m\bm{v}}-I_{m V^2/2}
=\frac{d}{2}n T \xi+{\sf P}^c:\nabla \bm{u}+\nabla\cdot\bm{q}^c ,
\end{equation}
where use has been made of Eqs.~\eqref{I_1=0} and \eqref{Garzo1.198}.

In the case of $\psi(\bm{v})=m v_\alpha v_\beta$, Eq.~\eqref{Garzo1.195} gives the relation
\begin{equation}\label{IVV}
\Lambda_{\alpha\beta}^E= -I_{m v_\alpha v_\beta}
+u_\alpha I_{m v_\beta}+u_\beta I_{mv_\alpha},
\end{equation}
where
\begin{eqnarray}
I_{mv_{\alpha}v_{\beta}}
&=&
\frac{m \sigma^{d-1}}{2}\int d\bm{v}_1\int d\bm{v}_2\int d\hat{\bm{\sigma}}
\Theta(\hat{\bm{\sigma}}\cdot\bm{v}_{12})
(\hat{\bm{\sigma}}\cdot\bm{v}_{12})
\{
[
v_{1,\alpha}'v_{1,\beta}'+v_{2,\alpha}'v_{2,\beta}'-v_{1,\alpha}v_{1,\beta}-v_{2,\alpha}v_{2,\beta}]
f^{(2)}(\bm{r},\bm{v}_1,\bm{r}+\bm{\sigma},\bm{v}_2;t)
\nonumber\\
&&
\quad
-\nabla \cdot \bm{\sigma} [v_{1,\alpha}'v_{1,\beta}'-v_{1,\alpha}v_{1,\beta}]
\int_0^1dx f^{(2)}(\bm{r}-x \bm{\sigma},\bm{v}_1,\bm{r}+(1-x)\bm{\sigma},\bm{v}_2;t) \} .
\end{eqnarray}
With the aid of
\begin{equation}\label{V_1+V_2}
v_{1,\alpha}'v_{1,\beta}'+v_{2,\alpha}'v_{2,\beta}'
-v_{1,\alpha}v_{1,\beta}-v_{2,\alpha}v_{2,\beta}
=-\frac{1+e}{2}
(\bm{v}_{12}\cdot\hat{\bm{\sigma}})
(v_{12,\alpha}\hat{\sigma}_\beta+\hat{\sigma}_\alpha v_{12,\beta})
+\frac{(1+e)^2}{2}(\bm{v}_{12}\cdot\hat{\bm{\sigma}})^2\hat{\sigma}_\alpha\hat{\sigma}_\beta,
\end{equation}
and
\begin{equation}\label{V_1V_1}
v_{1,\alpha}'v_{1,\beta}'-v_{1,\alpha}v_{1,\beta}
=-\frac{1+e}{2}(\hat{\bm{\sigma}}\cdot\bm{v}_{12})(\hat{\sigma}_\alpha v_{1,\beta}+\hat{\sigma}_\beta v_{1,\alpha})+\frac{(1+e)^2}{4}(\hat{\bm{\sigma}}\cdot\bm{v}_{12})^2\delta_{\alpha\beta},
\end{equation}
one achieves the result
\begin{equation}\label{lambda^E}
I_{mv_\alpha v_\beta}
=-\overline{\Lambda}_{\alpha\beta}^E-\nabla_\gamma \Sigma_{\alpha\beta\gamma},
\end{equation}
where
\begin{eqnarray}\label{over_Lambda}
\overline{\Lambda}_{\alpha\beta}^E
&=&
\frac{1+e}{4}m\sigma^{d-1}\int d\bm{v}_1 \int d\bm{v}_2 \int d\hat{\bm{\sigma}}
\Theta(\hat{\bm{\sigma}}\cdot\bm{v}_{12})(\hat{\bm{\sigma}}\cdot\bm{v}_{12})^2
\left\{
(v_{12,\alpha}\hat{\sigma}_\beta+\hat{\sigma}_\alpha v_{12,\beta})
f^{(2)}(\bm{r},\bm{v}_1,\bm{r}+\bm{\sigma},\bm{v}_2;t)\right.
\nonumber\\
&&\left.- (1+e)(\hat{\bm{\sigma}}\cdot\bm{v}_{12})
\hat{\sigma}_\alpha \hat{\sigma}_\beta
f^{(2)}(\bm{r},\bm{v}_1,\bm{r}+\bm{\sigma},\bm{v}_2;t)\right\},
\end{eqnarray}
and
\begin{equation}\label{Sigma}
\Sigma_{\alpha\beta\gamma}=\Sigma_{\alpha\beta\gamma}^{(1)}+
\delta_{\alpha\beta}S_\gamma
=\Sigma_{\alpha\beta\gamma}^{(1)}.
\end{equation}
The expression of $\Sigma_{\alpha\beta\gamma}^{(1)}$ is
\begin{eqnarray}\label{Sigma^1}
\Sigma_{\alpha\beta\gamma}^{(1)}
&=&\frac{1+e}{4}m\sigma^{d}\int d\bm{v}_1 \int d\bm{v}_2
\int d\hat{\bm{\sigma}}
\Theta(\hat{\bm{\sigma}}\cdot\bm{v}_{12})
(\hat{\bm{\sigma}}\cdot\bm{v}_{12})^2\hat{\sigma}_\gamma
(\hat{\sigma}_\alpha v_{1,\beta}+\hat{\sigma}_\beta v_{1,\alpha})
\int_0^1dx
f^{(2)}(\bm{r}-x \bm{\sigma},\bm{v}_1,\bm{r}+(1-x)\bm{\sigma},\bm{v}_2;t)
\nonumber\\
&=&
\frac{1+e}{4}m\sigma^{d}\int d\bm{v}_1 \int d\bm{v}_2
\int d\hat{\bm{\sigma}}
\Theta(\hat{\bm{\sigma}}\cdot\bm{v}_{12})
(\hat{\bm{\sigma}}\cdot\bm{v}_{12})^2\hat{\sigma}_\gamma
\left\{
\hat{\sigma}_\alpha (u_\beta+V_{G,\beta}+v_{12,\beta})+
\hat{\sigma}_\beta (u_\alpha+V_{G,\alpha}+v_{12,\alpha})
\right\}
\nonumber\\
&& \times
\int_0^1dx
f^{(2)}(\bm{r}-x \bm{\sigma},\bm{v}_1,\bm{r}+(1-x)\bm{\sigma},\bm{v}_2;t)
\nonumber\\
&=&
u_\alpha P^c_{\beta\gamma}+u_\beta P^c_{\alpha\gamma}+Q_{\alpha\beta\gamma}+\Upsilon_{\alpha\beta\gamma},
\end{eqnarray}
here we have introduced the quantities $Q_{\alpha\beta\gamma}$ and $\Upsilon_{\alpha\beta\gamma}$ as
\begin{eqnarray}\label{Q}
Q_{\alpha\beta\gamma}&=&
\frac{1+e}{4}m\sigma^{d}\int d\bm{v}_1 \int d\bm{v}_2
\int d\hat{\bm{\sigma}}
\Theta(\hat{\bm{\sigma}}\cdot\bm{v}_{12})
(\hat{\bm{\sigma}}\cdot\bm{v}_{12})^2
V_{G,\alpha}\hat{\sigma}_\beta \hat{\sigma}_\gamma
\int_0^1dx
f^{(2)}(\bm{r}-x \bm{\sigma},\bm{v}_1,\bm{r}+(1-x)\bm{\sigma},\bm{v}_2;t)
,
\\
\Upsilon_{\alpha\beta\gamma}&=&
\frac{1+e}{4}m\sigma^{d}\int d\bm{v}_1 \int d\bm{v}_2
\int d\hat{\bm{\sigma}}
\Theta(\hat{\bm{\sigma}}\cdot\bm{v}_{12})
(\hat{\bm{\sigma}}\cdot\bm{v}_{12})^2
v_{12,\alpha}\hat{\sigma}_\beta \hat{\sigma}_\gamma
\int_0^1dx
f^{(2)}(\bm{r}-x \bm{\sigma},\bm{v}_1,\bm{r}+(1-x)\bm{\sigma},\bm{v}_2;t)
\nonumber\\
&=&-
\frac{1+e}{4}m\sigma^{d}\int d\bm{v}_1 \int d\bm{v}_2
\int d\hat{\bm{\sigma}}
\Theta(\hat{\bm{\sigma}}\cdot\bm{v}_{12})
(\hat{\bm{\sigma}}\cdot\bm{v}_{12})^2
v_{12,\alpha}\hat{\sigma}_\beta \hat{\sigma}_\gamma
\int_0^1dx
f^{(2)}(\bm{r}-x \bm{\sigma},\bm{v}_1,\bm{r}+(1-x)\bm{\sigma},\bm{v}_2;t)\nonumber\\
&=& -\Upsilon_{\alpha\beta\gamma}=0.
\label{Upsilon}
\end{eqnarray}
As before, we have exchanged $1\leftrightarrow 2$ and have made the change of variable $\bm{\sigma} \to -\bm{\sigma}$ in the expression of $\Upsilon_{\alpha\beta\gamma}$. The quantity $Q_{\alpha\beta\gamma}$ satisfies the relation
\begin{equation}
q^c_\alpha=Q_{\alpha\beta\beta}.
\end{equation}
From Eqs.~\eqref{Garzo1.198}, \eqref{IVV}, \eqref{lambda^E} and \eqref{Sigma^1} we can rewrite $\Lambda_{\alpha\beta}^E$ as
\begin{eqnarray}\label{total_Lambda_E}
\Lambda_{\alpha\beta}^E&=&
\overline{\Lambda}_{\alpha\beta}^E+\nabla_\gamma \Sigma_{\alpha\beta\gamma}
+u_\alpha I_{m v_\beta}+u_\beta I_{mv_\alpha}
\nonumber\\
&=&\overline{\Lambda}_{\alpha\beta}^E+P_{\beta\gamma}^c\nabla_\gamma u_\alpha +P_{\alpha\gamma}^c\nabla_\gamma u_\beta+\nabla_\gamma Q_{\alpha\beta\gamma}
\nonumber\\
&=&\overline{\Lambda}_{\alpha\beta}^E+
\dot\gamma(\delta_{\alpha x}P_{\beta y}^c+\delta_{\beta x}P_{\alpha y}^c)
+\nabla_\gamma Q_{\alpha\beta\gamma}.
\end{eqnarray}
where use has been made of Eq.~\eqref{plane_shear} for the last identity in Eq.\ \eqref{total_Lambda_E}.

\section{Evaluation of $\Lambda^E_{\alpha\beta}$}\label{derivation_Lambda}

In this Appendix, we evaluate $\Lambda^E_{\alpha\beta}$ introduced in Eq.~\eqref{Garzo32} with the aid of Eqs.~\eqref{total_Lambda_E} and \eqref{Grad} under Enskog's approximation \eqref{decoupling}.

\subsection{Evaluation of $\overline{\Lambda}^E_{\alpha\beta}$}

The collisional moment $\overline{\Lambda}_{\alpha\beta}^E$ defined in Eq.~\eqref{total_Lambda_E} can be rewritten as
\begin{equation}
\label{Lambda_E}
\overline{\Lambda}_{\alpha\beta}^E=g_0 \Lambda_{\alpha\beta}^{(0)}+
\dot\gamma{\cal K}_y \left[\frac{\partial f}{\partial V_x} \right]_{\alpha\beta} ,
\end{equation}
where
\begin{eqnarray}
\Lambda_{\alpha\beta}^{(0)}
&=&-m \sigma^{d-1} \int d\bm{v}_1\int d\bm{v}_2\int d\hat{\bm{\sigma}}\Theta(\bm{v}_{12}\cdot\hat{\bm{\sigma}})
|\bm{v}_{12}\cdot \hat{\bm{\sigma}}|V_{1,\alpha} V_{1,\beta}
\left\{
\frac{f(\bm{V}_1^{''})f(\bm{V}_2^{''})}{e^2}-f(\bm{V}_1)f(\bm{V}_2)
\right\}
\nonumber\\
&=&
-\frac{m  \sigma^{d-1} }{2}
\int d\bm{v}_1\int d\bm{v}_2\int d\hat{\bm{\sigma}}\Theta(\bm{v}_{12}\cdot\hat{\bm{\sigma}})(\bm{v}_{12}\cdot\hat{\bm{\sigma}})
f(\bm{V}_1)f(\bm{V}_2)\nonumber\\
& & \times (V_{1,\alpha}'V_{1,\beta}'+V_{2,\alpha}'V_{2,\beta}'
-V_{1,\alpha}V_{1,\beta}-V_{2,\alpha}V_{2,\beta}),
\end{eqnarray}
and
\begin{equation}
{\cal K}_\mu [X_\nu]_{\alpha\beta}
=-g_0 m \sigma^d
\int d\bm{v}_1 \int d\bm{v}_2 \int d\hat{\bm{\sigma}} \Theta(\hat{\bm{\sigma}} \cdot \bm{v}_{12})
(\hat{\bm{\sigma}}\cdot\bm{v}_{12})\hat{\sigma}_\mu V_{1,\alpha}V_{1,\beta}
\left[ \frac{f(\bm{V}_1^{''})X_\nu (\bm{V}_2^{''})}{e^2}+f(\bm{V}_1)X_\mu(\bm{V}_2) \right].
\end{equation}

When we adopt Eq.~\eqref{Grad} in the expression of $\Lambda_{\alpha\beta}^{(0)}$ and neglect quadratic contributions in the stress tensor, one gets the result \cite{Chamorro15,Garzo02}
\begin{equation}
\label{Lambda_maintext}
\Lambda_{\alpha\beta}^{(0)}=n T\left(
\nu \Pi_{\alpha\beta}+\lambda\delta_{\alpha\beta}\right),
\end{equation}
where $\nu$ and $\lambda$ are given by Eqs.\ \eqref{nu} and \eqref{nu'}, respectively.

Similarly, the collisional moment ${\cal K}_y\left[\partial_{V_x} f\right]_{\alpha\beta}$ is
\begin{eqnarray}\label{1st_K}
{\cal K}_y\left[\frac{\partial f}{\partial V_x}\right]_{\alpha\beta}
&=& -m\sigma^d g_0 \int d\bm{V}_1 \int d\bm{V}_2\int d\hat{\bm{\sigma}}
\Theta(\hat{\bm{\sigma}}\cdot\bm{v}_{12})(\hat{\bm{\sigma}}\cdot\bm{v}_{12})\hat{\sigma}_y V_{1,\alpha}V_{1,\beta}
\left[e^{-2} f(\bm{V}_1^{''})\frac{\partial f(\bm{V}_2^{''})}{\partial V_{2,x}}
+f(\bm{V}_1)\frac{\partial f(\bm{V}_2)}{\partial V_{2,x}} \right] \nonumber\\
&=&
m\sigma^d g_0 \int d\bm{V}_1 \int d\bm{V}_2
\int d\bm{V}_2\int d\hat{\bm{\sigma}}
\Theta(\hat{\bm{\sigma}}\cdot\bm{v}_{12})(\hat{\bm{\sigma}}\cdot\bm{v}_{12})\hat{\sigma}_y
f(\bm{V}_1)\frac{\partial f(\bm{V}_2)}{\partial V_{2,x}}
(V_{1,\alpha}'V_{2,\beta}'-V_{1,\alpha}V_{2,\beta}) ,
\end{eqnarray}
where we have used the post-collisional velocities $\bm{V}_i'$. Taking into account the relation
\begin{equation}
V_{1,\alpha}'V_{1,\beta}'=
V_{1,\alpha}V_{2,\beta}
-\frac{1+e}{2}(\hat{\bm{\sigma}}\cdot\bm{v}_{12})(\hat{\sigma}_\alpha V_{1,\beta}+\hat{\sigma}_\beta V_{1,\alpha})+
\frac{(1+e)^2}{4}(\hat{\bm{\sigma}}\cdot\bm{v}_{12})^2\hat{\sigma}_\alpha \hat{\sigma}_\beta
,
\end{equation}
Eq.~\eqref{1st_K} can be rewritten as
\begin{eqnarray}
\label{2nd_K}
{\cal K}_y\left[\frac{\partial f}{\partial V_x}\right]_{\alpha\beta}
&=& -g_0 m\sigma^d \frac{1+e}{4}
\int d\bm{V}_1 \int d\bm{V}_2
f(\bm{V}_1)\frac{\partial f(\bm{V}_2)}{\partial V_{2,x}}
\int d\hat{\bm{\sigma}}
\Theta(\hat{\bm{\sigma}}\cdot\bm{v}_{12})(\hat{\bm{\sigma}}\cdot\bm{v}_{12})^2
\hat{\sigma}_y \nonumber\\
& & \times
[2(V_{1,\beta}\hat{\sigma}_\alpha +V_{1,\alpha}\hat{\sigma}_\beta
-(1+e) (\hat{\bm{\sigma}}\cdot\bm{v}_{12})\hat{\sigma}_\alpha\hat{\sigma}_\beta]
\nonumber\\
&=& -g_0 m\sigma^d \frac{1+e}{4}\int d\bm{V}_1 \int d\bm{V}_2 f(\bm{V}_1)f(\bm{V}_2)
\int d\hat{\bm{\sigma}}
\Theta(\hat{\bm{\sigma}}\cdot\bm{v}_{12})(\hat{\bm{\sigma}}\cdot\bm{v}_{12})
\hat{\sigma}_x\hat{\sigma}_y
\nonumber\\
&& \times
[4(V_{1,\alpha}\hat{\sigma}_\beta+V_{1,\beta}\hat{\sigma}_\alpha)-
3(1+e)(\hat{\bm{\sigma}}\cdot\bm{v}_{12})\hat{\sigma}_\alpha\hat{\sigma}_\beta] ,
\end{eqnarray}
where we have used the integral by parts and
$\partial_{V_{2,x}}
(\hat{\bm{\sigma}}\cdot\bm{V}_{12})^n=-n(\hat{\bm{\sigma}}\cdot\bm{V}_{12})^{n-1}\hat{\sigma}_x$. Equation \eqref{2nd_K} can be expressed in a more compact form as
\begin{equation}
\label{2ndK_yf_x}
{\cal K}_y\left[\frac{\partial f}{\partial V_x}\right]_{\alpha\beta}
=
-g_0 m\sigma^d \frac{1+e}{4}\int d\bm{V}_1 \int d\bm{V}_2 f(\bm{V}_1)f(\bm{V}_2)
\{ 4(V_{1,\alpha}\Xi_\beta+V_{1,\beta} \Xi_\alpha)-3(1+e) \Omega_{\alpha\beta} \} ,
\end{equation}
where
\begin{eqnarray}
\label{Xi}
\Xi_\alpha&=& \int d\hat{\bm{\sigma}} \Theta(\hat{\bm{\sigma}}\cdot\bm{v}_{12})(\hat{\bm{\sigma}}\cdot\bm{v}_{12})
\hat{\sigma}_x\hat{\sigma}_y \hat{\sigma}_\alpha
=\frac{B_2}{d+2}(\delta_{\alpha x}V_{12,y}+\delta_{\alpha y}V_{12,x}),
\\
\Omega_{\alpha\beta}&=&
\int d\hat{\bm{\sigma}}\Theta(\hat{\bm{\sigma}}\cdot\bm{v}_{12})(\hat{\bm{\sigma}}\cdot\bm{v}_{12})^2
\hat{\sigma}_x\hat{\sigma}_y \hat{\sigma}_\alpha\hat{\sigma}_\beta
\nonumber\\
&=&\frac{B_2}{(d+2)(d+4)}[2(V_{12,\alpha}V_{12,x}\delta_{\beta y}+V_{12,\alpha}V_{12,y}\delta_{\beta x}+
V_{12,\beta}V_{12,x}\delta_{\alpha y}+V_{12,\beta}V_{12,y}\delta_{\alpha x}+ V_{12,x}V_{12,y}\delta_{\alpha\beta})
\nonumber\\
&&
+V_{12}^2(\delta_{\alpha x}\delta_{\beta y}+\delta_{\alpha y}\delta_{\beta x})] .
\end{eqnarray}

In order to evaluate Eq.\ \eqref{2ndK_yf_x}, Grad's distribution function \eqref{Grad} is considered. This distribution can be decomposed in the form
\begin{equation}\label{Grad_expand}
f(\bm{V})=f_{\rm M}(\bm{V})+f^{(1)}(\bm{V}), \quad
f^{(1)}(\bm{V})=\frac{m}{2T} f_{\rm M}(\bm{V})\Pi_{\alpha\beta} V_\alpha V_\beta .
\end{equation}
When one replaces $f$ by its Grad's approximation \eqref{Grad_expand} in Eq.~\eqref{2ndK_yf_x}, the integral on the right hand side of Eq.~\eqref{2ndK_yf_x} consists of the following two contributions. The first contribution is
\begin{equation}
\int d\bm{V}_1 \int d\bm{V}_2 f_{\rm M}(\bm{V}_1)f_{\rm M}(\bm{V}_2)
(V_{1,\alpha}\Xi_\beta+V_{1,\beta}\Xi_\alpha)=
\frac{2n^2TB_2}{(d+2)m} I_{\alpha\beta},
\end{equation}
where
\begin{eqnarray}
I_{\alpha\beta}&=&\frac{1}{\pi^d}\int d\bm{G} \int d\bm{g}e^{-2G^2-\frac{g^2}{2}}
\left[
\left(G_\alpha+\frac{g_\alpha}{2}\right)
\left\{\delta_{\beta x}g_y+\delta_{\beta y}g_x \right\}
+\left(G_\beta+\frac{g_\beta}{2}\right)
\left\{\delta_{\alpha x}g_y+\delta_{\alpha y}g_x
 \right\}
 \right] \nonumber\\
&=& \delta_{\alpha x}\delta_{\beta y}+\delta_{\alpha y}\delta_{\beta x},
\end{eqnarray}
with $\bm{G}\equiv \sqrt{m/(2T)}(\bm{V}_1+\bm{V}_2)/2$ and
$\bm{g}\equiv \sqrt{m/(2T)}(\bm{V}_1-\bm{V}_2)$.
Therefore, we obtain
\begin{equation}\label{int_I}
\int d\bm{V}_1 \int d\bm{V}_2 f_{\rm M}(\bm{V}_1)f_{\rm M}(\bm{V}_2)
(V_{1,\alpha}\Xi_\beta+V_{1,\beta}\Xi_\alpha)=\frac{2n^2TB_2}{(d+2)m}
(\delta_{\alpha x}\delta_{\beta y}+\delta_{\alpha y}\delta_{\beta x}) .
\end{equation}

The second contribution is given by
\begin{equation}\label{int_Omega}
\int d\bm{V}_1 \int d\bm{V}_2 f_{\rm M}(\bm{V}_1)f_{\rm M}(\bm{V}_2) \Omega_{\alpha\beta}
=\frac{2n^2 T B_2}{(d+2)(d+4)m}
\int d\bm{G} \int d\bm{g} \frac{e^{-2G^2-g^2/2}}{\pi^d}
\tilde{\Omega}_{\alpha\beta}
=\frac{2^{1-d/2}n^2 T B_2}{(d+2)(d+4)m}
 \int d\bm{g} \frac{e^{-g^2/2}}{\pi^{d/2}}
\tilde{\Omega}_{\alpha\beta} ,
\end{equation}
where $\tilde{\Omega}_{\alpha\beta}=2(g_\alpha g_x \delta_{\beta y}+g_\alpha g_y \delta_{\beta x}
+g_\beta g_x \delta_{\alpha y}+g_\beta g_y \delta_{\alpha x}+ g_x g_y \delta_{\alpha \beta})
+g^2(\delta_{\alpha x}\delta_{\beta y}+\delta_{\alpha y}\delta_{\beta x})$.
The integration over $\mathbf{g}$ in Eq.\ \eqref{int_Omega} gives the result
\begin{eqnarray}
\int d\bm{g} \frac{e^{-g^2/2}}{\pi^{d/2}}
\tilde{\Omega}_{\alpha\beta}
&=&
(\delta_{\alpha x}\delta_{\beta y}+\delta_{\alpha y}\delta_{\beta x})
\int d\bm{g} \frac{e^{-g^2/2}}{\pi^{d/2}}\left[2(g_x^2+g_y^2)+g^2 \right]
=2^{d/2}(d+4)(\delta_{\alpha x}\delta_{\beta y}+\delta_{\alpha y}\delta_{\beta x}).
\end{eqnarray}
Therefore, we obtain
\begin{equation}\label{Omega_int_result}
\int d\bm{V}_1 \int d\bm{V}_2 f_{\rm M}(\bm{V}_1)f_{\rm M}(\bm{V}_2) \Omega_{\alpha\beta}
=\frac{2n^2T B_2}{(d+2)m}(\delta_{\alpha x}\delta_{\beta y}+\delta_{\alpha y}\delta_{\beta x}) .
\end{equation}
Substituting Eqs.~\eqref{int_I} and \eqref{Omega_int_result} with $f=f_{\rm M}$ into Eq.~\eqref{2nd_K} yields
\begin{eqnarray}\label{K_y_f_eq}
{\cal K}_y\left[\frac{\partial f_{\rm M}}{\partial V_x}\right]_{\alpha\beta}
&=&
-(\delta_{\alpha x}\delta_{\beta y}+\delta_{\alpha y}\delta_{\beta x})
\frac{2n^2T B_2}{4(d+2)}
g_0  \sigma^d(1+e)(1-3e) \nonumber\\
&=&
 -\frac{2^{d-2}}{d+2}nT \varphi g_0 (1+e) (1-3e)
(\delta_{\alpha x}\delta_{\beta y}+\delta_{\alpha y}\delta_{\beta x}) ,
\end{eqnarray}
where use has been made of Eqs.~\eqref{volume_fraction}, \eqref{beta}, \eqref{PoF_B14}, and \eqref{PoF_B15}.

Similarly, the contribution coming from $f^{(1)}$ in Eq.~\eqref{2ndK_yf_x} can be evaluated as
\begin{eqnarray}\label{K_y_f_1}
{\cal K}_y\left[\frac{\partial f^{(1)}}{\partial V_x}\right]_{\alpha\beta}
&=&
-g_0 m^2 \sigma^d\frac{(1+e)}{8T}\Pi_{\mu\nu}
\int d\bm{V}_1 \int d\bm{V}_2 f_{\rm M}(\bm{V}_1)f_{\rm M}(\bm{V}_2)
 (V_{1,\mu}V_{1,\nu}+V_{2,\mu}V_{2,\nu} )
\nonumber\\
&& \times
\{ 4(V_{1,\alpha}\Xi_\beta+V_{1,\beta} \Xi_\alpha)-3(1+e) \Omega_{\alpha\beta} \}
\nonumber\\
&=& -g_0 m^2 \sigma^d\frac{(1+e)}{8T}(4{\cal A}_{\alpha\beta}-3(1+e){\cal B}_{\alpha\beta}).
\end{eqnarray}
The first contribution ${\cal A}_{\alpha\beta}$ is given by
\begin{equation}
{\cal A}_{\alpha\beta}\equiv \Pi_{\mu\nu}\int d\bm{V}_1 \int d\bm{V}_2 f_{\rm M}(\bm{V}_1)f_{\rm M}(\bm{V}_2)
(V_{1,\mu}V_{1,\nu}+V_{2,\mu}V_{2,\nu})
(V_{1,\alpha}\Xi_\beta+V_{1,\beta}\Xi_\alpha)=
\frac{2n^2B_2}{d+2} \left(\frac{2T}{m}\right)^{2} \Pi_{\mu\nu}J_{\alpha\beta\mu\nu},
\end{equation}
where we have introduced
\begin{eqnarray}
J_{\alpha\beta\mu\nu}&=&\frac{1}{\pi^d}\int d\bm{G} \int d\bm{g}e^{-2G^2-\frac{g^2}{2}}
\left(G_\mu G_\nu+\frac{g_\mu g_\nu}{4} \right)
\left[
\left(G_\alpha+\frac{g_\alpha}{2}\right)
\left\{\delta_{\beta x}g_y+\delta_{\beta y}g_x \right\}
+\left(G_\beta+\frac{g_\beta}{2}\right)
\left\{\delta_{\alpha x}g_y+\delta_{\alpha y}g_x
 \right\}
 \right]
 \nonumber\\
&=& J_{\alpha\beta\mu\nu}^{(1)}+J_{\alpha\beta\mu\nu}^{(2)}
\end{eqnarray}
with
\begin{eqnarray}
J_{\alpha\beta\mu\nu}^{(1)}
&=& \frac{1}{2\pi^d}\int d\bm{G} \int d\bm{g} e^{-2G^2-g^2/2}
G_\mu G_\nu(g_\alpha g_y \delta_{\beta x}+g_\alpha g_x\delta_{\beta y}+g_\beta g_y \delta_{\alpha x}+g_x g_\beta \delta_{\alpha y}) , \\
J_{\alpha\beta\mu\nu}^{(2)}
&=& \frac{1}{8\pi^d} \int d\bm{G} \int d\bm{g} e^{-2G^2-g^2/2}
g_\mu g_\nu (g_\alpha g_y \delta_{\beta x}+g_\alpha g_x\delta_{\beta y}+g_\beta g_y \delta_{\alpha x}+g_x g_\beta \delta_{\alpha y} ) .
\end{eqnarray}
Here, it is straightforward to show that
\begin{equation}
\Pi^k_{\mu\nu}J^{(1)}_{\alpha\beta\mu\nu}=0,
\end{equation}
because of $\Pi_{\mu\nu}\int d\bm{G} e^{-2G^2}G_\mu G_\nu\propto \Pi^k_{\mu\nu}\delta_{\mu\nu}=\Pi^k_{\mu\mu}=0$.
On the other hand, we have the relation
\begin{eqnarray}
\Pi^k_{\mu\nu}J_{\alpha\beta\mu\nu}^{(2)}
&=&
\frac{1}{4}[\Pi^k_{\alpha x}\delta_{\beta y}+\Pi^k_{\alpha y}\delta_{\beta x}+
 \Pi^k_{\beta x}\delta_{\alpha y}+\Pi^k_{\beta y}\delta_{\alpha x}] ,
\end{eqnarray}
where we have taken into account the intermediate result
\begin{equation}
\int d\bm{g} e^{-g^2/2} g_\mu g_\nu g_\alpha g_\beta
=\frac{S_d}{d(d+2)}\int_0^\infty dg g^{d+3}e^{-g^2/2}(\delta_{\mu\nu}\delta_{\alpha\beta}+\delta_{\alpha\mu}\delta_{\beta\nu}+
\delta_{\alpha\nu}\delta_{\beta\mu})
=2^{d/2}\pi^{d/2}(\delta_{\mu\nu}\delta_{\alpha\beta}+\delta_{\alpha\mu}\delta_{\beta\nu}+
\delta_{\alpha\nu}\delta_{\beta\mu}) .
\end{equation}
Here, $S_d=2\pi^{d/2}/\Gamma(d/2)$ is the total solid angle in $d$ dimensions. Therefore, we obtain
\begin{equation}
{\cal A}_{\alpha\beta}
=
\frac{n^2B_2}{2(d+2)}\left(\frac{2T}{m}\right)^{2}
[\Pi^k_{\alpha x}\delta_{\beta y}+\Pi^k_{\alpha y}\delta_{\beta x}+
 \Pi^k_{\beta x}\delta_{\alpha y}+\Pi^k_{\beta y}\delta_{\alpha x}] .
\label{1stK_yf_1_Vx}
\end{equation}

The second contribution ${\cal B}_{\alpha\beta}$ in Eq.~\eqref{K_y_f_1} is given by
\begin{eqnarray}\label{PiOmega}
{\cal B}_{\alpha\beta}&\equiv&\Pi_{\mu\nu}\int d\bm{V}_1\int d\bm{V}_2f_{\rm M}(\bm{V}_1)f_{\rm M}(\bm{V}_2)
(V_{1,\mu}V_{1,\nu}+V_{2,\mu}V_{2,\nu})\Omega_{\alpha\beta}
\nonumber\\
&=&
4\frac{2^{-1-d/2}n^2 B_2}{d(d+2)^2(d+4)}
\left(\frac{2T}{m}\right)^{2}
 \int \frac{d\bm{g}}{\pi^{d/2}}e^{-g^2/2}g^4
[\Pi^k_{\alpha x}\delta_{\beta y}+\Pi^k_{\alpha y}\delta_{\beta x}
+\Pi^k_{\beta x}\delta_{\alpha y}+\Pi^k_{\beta y}\delta_{\alpha x}+\Pi^k_{xy}\delta_{\alpha\beta}]
\nonumber\\
&=&
\frac{2 n^2 B_2}{(d+2)(d+4)}
\left(\frac{2T}{m}\right)^{2}
[\Pi^k_{\alpha x}\delta_{\beta y}+\Pi^k_{\alpha y}\delta_{\beta x}
+\Pi^k_{\beta x}\delta_{\alpha y}+\Pi^k_{\beta y}\delta_{\alpha x}+\Pi^k_{xy}\delta_{\alpha\beta}] .
\end{eqnarray}

From Eqs.~\eqref{1stK_yf_1_Vx} and \eqref{PiOmega} one achieves
\begin{equation}
\label{vic0}
4{\cal A}_{\alpha\beta}-3(1+e){\cal B}_{\alpha\beta}
=\frac{2n^2B_2}{(d+2)(d+4)}\left(\frac{2T}{m}\right)^2
\{
(d+1-3e)
[\Pi^k_{\alpha x}\delta_{\beta y}+\Pi^k_{\alpha y}\delta_{\beta x}+\Pi^k_{\beta x}\delta_{\alpha y}+\Pi^k_{\beta y}\delta_{\alpha x}]
-3(1+e)\delta_{\alpha\beta} \Pi^k_{xy}
\} .
\end{equation}
The final expression for ${\cal K}_y[\partial_{V_x}f^{(1)}]_{\alpha\beta}$ is obtained after substituting Eq.\ \eqref{vic0} into Eq.~\eqref{K_y_f_1}. The result is
\begin{eqnarray}\label{Kyf1:result}
{\cal K}_y\left[\frac{\partial f^{(1)}}{\partial V_x}\right]_{\alpha\beta}
&=&
-\frac{g_0 (1+e) n^2\sigma^d T B_2}{(d+2)(d+4)}
\{
(d+1-3e)
[\Pi^k_{\alpha x}\delta_{\beta y}+\Pi^k_{\alpha y}\delta_{\beta x}+\Pi^k_{\beta x}\delta_{\alpha y}+\Pi^k_{\beta y}\delta_{\alpha x}]
-3(1+e)\delta_{\alpha\beta} \Pi^k_{xy}
\}
\nonumber\\
&=&
-\frac{2^{d-1}}{(d+2)(d+4)}
n T \varphi g_0(1+e)
\{
(d+1-3e)(\Pi^k_{\alpha x}\delta_{\beta y}+\Pi^k_{\alpha y}\delta_{\beta x}
+\Pi^k_{\beta x}\delta_{\alpha y}+\Pi^k_{\beta y}\delta_{\alpha x})
\nonumber\\
& &
-
3(1+e)\delta_{\alpha\beta}\Pi^k_{xy}
\} .
\end{eqnarray}
Equation \eqref{Lambda_E:result} is easily obtained by substituting Eqs.~\eqref{Lambda_maintext}, \eqref{K_y_f_eq} and \eqref{Kyf1:result} into Eq.~\eqref{Lambda_E}.

\subsection{Evaluation of $Q_{\alpha\beta\gamma}$}

In this subsection, the quantity $Q_{\alpha\beta\gamma}$ introduced in Eq.~\eqref{Q} is determined by using Grad's approximation \eqref{Grad}. According to the symmetry of the simple shear flow, it is expected that $Q_{\alpha\beta\gamma}=0$. Substitution of Eqs.~\eqref{int_f_2} and \eqref{Garzo_A17} into Eq.~\eqref{Q} leads to
\begin{eqnarray}
\label{Q_1}
Q_{\alpha\beta\gamma}
&\approx& \frac{1+e}{4}g_0 m\sigma^d\int d\bm{v}_1 \int d\bm{v}_2 \int d\hat{\bm{\sigma}}\Theta(\hat{\bm{\sigma}}\cdot\bm{v}_{12})(\hat{\bm{\sigma}}\cdot\bm{v}_{12})^2 V_{G,\alpha}\hat{\sigma}_\alpha\hat{\sigma}_\gamma
f\left(\bm{V}_1+\frac{1}{2}\dot\gamma \sigma \hat{\sigma}_y\bm{e}_x \right)
f\left(\bm{V}_2-\frac{1}{2}\dot\gamma \sigma \hat{\sigma}_y\bm{e}_x \right)
\nonumber\\
&=& \frac{1+e}{2}n^2\sigma^d g_0 T \displaystyle\sqrt{\frac{2T}{m}}\tilde{Q}_{\alpha\beta\gamma},
\end{eqnarray}
where
\begin{eqnarray}\label{Q=0}
\tilde{Q}_{\alpha\beta\gamma}
&=& \frac{1}{\pi^d}\int d\bm{G} \int d\bm{g} \int d\hat{\bm{\sigma}}
\Theta(\hat{\bm{\sigma}}\cdot\bm{g})(\hat{\bm{\sigma}}\cdot\bm{g})^2 G_\alpha
\hat{\sigma}_\beta \hat{\sigma}_\gamma
\exp\left[-\left\{2G^2+\frac{g^2}{2}+\dot\gamma'  \hat{\sigma}_y g_x+\frac{1}{2}\dot{\gamma'}^2\hat{\sigma}_y^2 \right\} \right]
\nonumber\\
&& \times
\left[
1+ \frac{m}{2T}\Pi^k_{\alpha\beta}
\left(
2G_\alpha G_\beta+\frac{g_\alpha g_\beta}{2}
\right)
\right]
\nonumber\\
&=&0.
\end{eqnarray}
Here, the parameter $\dot\gamma'\equiv \dot\gamma^*\tau_T$ has been introduced. Thus, we immediately conclude that the collisional contribution to the heat flux vanishes, namely,
\begin{equation}\label{q_c=0}
\bm{q}^c=\bm{0} .
\end{equation}
In summary, the contributions of $Q_{\alpha\beta\gamma}$ and $\bm{q}^c$ become zero if we adopt Eq.~\eqref{Grad} for the velocity distribution function.
This is the expected result since $Q_{\alpha\beta\gamma}$ is related to the collisional contribution to the heat flux, which must be decoupled with the stress perturbation as in Eq.~\eqref{Grad}.

\section{Evaluation of the collisional stress}\label{collision_stress}

In this Appendix, the collisional stress $P_{\alpha\beta}^c$ given by Eq.~\eqref{P_c:main_text} is obtained
within the framework of Enskog's kinetic theory and Grad's approximation Eq.~\eqref{Grad}. The outline of this Appendix follows Ref.~\cite{Garzo13}.  Let us decompose first $P_{\alpha\beta}^c$ in two parts:
\begin{equation}
\label{collisional_stress_sum}
P_{\alpha\beta}^c=P_{\alpha\beta}^{c(0)}+P_{\alpha\beta}^{c(1)} ,
\end{equation}
where $P_{\alpha\beta}^{c(0)}$ and $P_{\alpha\beta}^{c(1)}$ are, respectively, given by
\begin{eqnarray}
\label{P_c0}
P_{\alpha\beta}^{c(0)}&=& \frac{(1+e)}{4} m\sigma^d g_0 \int d\bm{V}_1 \int d\bm{V}_2\int d\hat{\bm{\sigma}}  \Theta(\bm{v}_{12}\cdot\hat{\bm{\sigma}})
(\bm{v}_{12}\cdot\hat{\bm{\sigma}})^2\hat{\sigma}_\alpha \hat{\sigma}_\beta
f_{\rm M}\left(\bm{V}_1+\frac{1}{2}\dot\gamma \sigma \hat{\sigma}_y \bm{e}_x\right)
f_{\rm M}\left(\bm{V}_2-\frac{1}{2}\dot\gamma \sigma \hat{\sigma}_y \bm{e}_x \right),  \\
P_{\alpha\beta}^{c(1)}&=&
\frac{(1+e)}{8T} m^2\sigma^d g_0 \Pi_{\mu\nu} \int d\bm{V}_1 \int d\bm{V}_2\int d\hat{\bm{\sigma}}  \Theta(\bm{v}_{12}\cdot\hat{\bm{\sigma}})
(\bm{v}_{12}\cdot\hat{\bm{\sigma}})^2\hat{\sigma}_\alpha\hat{\sigma}_\beta
f_{\rm M}\left(\bm{V}_1+\frac{1}{2}\dot\gamma \sigma \hat{\sigma}_y \bm{e}_x\right)
f_{\rm M}\left(\bm{V}_2-\frac{1}{2}\dot\gamma \sigma \hat{\sigma}_y \bm{e}_x \right)
\nonumber\\
&&
\times
\left\{
\left(V_{1,\mu}+\frac{1}{2}\dot\gamma \sigma \hat{\sigma}_y\delta_{x\mu}\right)
\left(V_{1,\nu}-\frac{1}{2}\dot\gamma \sigma \hat{\sigma}_y\delta_{x\nu} \right)
+
\left(V_{2,\mu}+\frac{1}{2}\dot\gamma \sigma \hat{\sigma}_y\delta_{x\mu}\right)
\left(V_{2,\nu}-\frac{1}{2}\dot\gamma \sigma \hat{\sigma}_y\delta_{x\nu} \right)
\right\} .
\label{P_c1}
\end{eqnarray}

First, let us evaluate $P_{\alpha\beta}^{c(0)}$. This quantity can be rewritten in dimensionless form as
\begin{equation}\label{dim_P(0)}
P_{\alpha\beta}^{c(0)}
=\frac{1+e}{2}n^2\sigma^d\chi T \tilde{P}_{\alpha\beta}^{c(0)} ,
\end{equation}
where
\begin{eqnarray}\label{dimless_P(0)}
\tilde{P}_{\alpha\beta}^{c(0)}&\equiv&
\frac{1}{\pi^d}\int d\bm{G} \int d\bm{g}
\int d\hat{\bm{\sigma}}  \Theta(\bm{g}\cdot\hat{\bm{\sigma}})(\bm{g}\cdot\hat{\bm{\sigma}})^2\hat{\sigma}_\alpha \hat{\sigma}_\beta
\exp\left[-\left\{2G^2+\frac{g^2}{2}+\dot\gamma'  \hat{\sigma}_y g_x+\frac{1}{2}\dot{\gamma'}^2\hat{\sigma}_y^2 \right\} \right]
\nonumber\\
&=&
\frac{1}{(2\pi)^{d/2}}\int d\bm{g}\int d\hat{\bm{\sigma}}\Theta(\bm{g}\cdot\hat{\bm{\sigma}})(\bm{g}\cdot\hat{\bm{\sigma}})^2
\hat{\sigma}_\alpha \hat{\sigma}_\beta\exp\left[-\frac{g^2+2\dot\gamma'\hat{\sigma}_yg_x+\dot{\gamma'}^2\hat{\sigma}_y^2}{2}\right]
.
\end{eqnarray}
Because we cannot perform the angular integral of Eq.~\eqref{dimless_P(0)} we expand it as a series of powers of $\dot\gamma'$.
As shown in Figs.~\ref{fig_tauT} and \ref{fig_tauT_e_change}, the parameter $\gamma'=\dot\gamma^*\tau_T$ is small in the high shear regime  for not quite strong inelasticity. Therefore, we only keep linear terms in $\dot\gamma'$ in the evaluation of $P_{\alpha\beta}^{c(0)}$. The result is
\begin{equation}
\label{dimless_P(0)_bis}
P_{\alpha\beta}^{c(0)}\approx \frac{1}{(2\pi)^{d/2}}
\int d\bm{g}
\int d\hat{\bm{\sigma}}  \Theta(\bm{g}\cdot\hat{\bm{\sigma}})(\bm{g}\cdot\hat{\bm{\sigma}})^2\hat{\sigma}_\alpha \hat{\sigma}_\beta
e^{-g^2/2}\left[1-\dot\gamma' \hat{\sigma}_y g_x \right].
\end{equation}
Equation \eqref{dimless_P(0)_bis} can be rewritten as
\begin{equation}\label{P_c0}
\tilde{P}_{\alpha\beta}^{c(0)}\approx \tilde{P}_{\alpha\beta}^{c(0,0)}-
\dot\gamma' \tilde{P}_{\alpha\beta}^{c(0,1)},
\end{equation}
where
\begin{eqnarray}\label{P_c00}
\tilde{P}_{\alpha\beta}^{c(0,0)}
&=& \frac{1}{\pi^d}\int d\bm{G} \int d\bm{g}
\int d\hat{\bm{\sigma}}  \Theta(\bm{g}\cdot\hat{\bm{\sigma}})(\bm{g}\cdot\hat{\bm{\sigma}})^2\hat{\sigma}_\alpha
\hat{\sigma}_\beta e^{-2G^2-g^2/2}
=
\frac{\pi^{d/2}}{d\Gamma(d/2)}\delta_{\alpha\beta} ,
\end{eqnarray}
and
\begin{eqnarray}
\label{P_c01}
\tilde{P}_{\alpha\beta}^{c(0,1)}&=& \frac{1}{\pi^d}\int d\bm{G} \int d\bm{g}
\int d\hat{\bm{\sigma}}  \Theta(\bm{g}\cdot\hat{\bm{\sigma}})(\bm{g}\cdot\hat{\bm{\sigma}})^2 \hat{\sigma}_\alpha \hat{\sigma}_\beta \hat{\sigma}_y
g_x e^{-2G^2-g^2/2}
\nonumber\\
&=&
\frac{B_3}{2^{d/2}\pi^{d/2}(d+3)}
\int d\bm{g} e^{-g^2/2}
[ g^{-1} g_\alpha g_\beta g_x g_y+
g_x g(\delta_{\alpha\beta} g_y+\delta_{\beta y} g_\alpha+\delta_{y \alpha}g_\beta)]
\nonumber\\
&=&
\frac{2\sqrt{2}\pi^{(d-1)/2}}{d(d+2)\Gamma(d/2)}
\left\{\delta_{\alpha x}\delta_{\beta y}+\delta_{\alpha y}\delta_{\beta x} \right\}.
\end{eqnarray}
Substitution of Eqs.~\eqref{P_c00} and \eqref{P_c01} into Eq.~\eqref{P_c0} yields
\begin{equation}
\label{P_c(0):result}
\tilde{P}_{\alpha\beta}^{c(0)}
\approx
\frac{\pi^{d/2}}{d\Gamma(d/2)}
\left[ \delta_{\alpha\beta}
-\dot\gamma'
\frac{2\sqrt{2}}{\sqrt{\pi}(d+2)}
\left\{\delta_{\alpha x}\delta_{\beta y}+\delta_{\alpha y}\delta_{\beta x} \right\}
\right] .
\end{equation}

The contribution $P_{\alpha\beta}^{c(1)}$ can be rewritten as
\begin{equation}\label{P_c(1)}
P_{\alpha\beta}^{c(1)}=\frac{1+e}{2}g_0 n^2 \sigma^d T \tilde{P}_{\alpha\beta}^{c(1)} ,
\end{equation}
where
\begin{eqnarray}
\label{vic_pc}
\tilde{P}_{\alpha\beta}^{c(1)}&=&
\Pi^k_{\mu\nu} \int \frac{d\bm{G}}{\pi^{d/2}}\int \frac{d\bm{g}}{\pi^{d/2}}
\int d\hat{\bm{\sigma}}\Theta(\bm{g}\cdot\hat{\bm{\sigma}})
(\bm{g}\cdot\hat{\bm{\sigma}})^2
\hat{\sigma}_\alpha\hat{\sigma}_\beta
\exp
\left[-\left\{
2G^2+\frac{g^2}{2}+\dot\gamma'\hat{\sigma}_y g_x +\frac{1}{2}\dot\gamma'^2
\hat{\sigma}_y^2
+
\dot{\gamma'}^{2}\hat{\sigma}_y^2
\right\}
\right]
\nonumber\\
&& \times
 \left\{\left(G_\mu+\frac{g_\mu}{2}+\frac{1}{2}\dot\gamma'\hat{\sigma}_y\delta_{x\mu} \right)
\left(G_\nu+\frac{g_\nu}{2}+\frac{1}{2}\dot\gamma'\hat{\sigma}_y\delta_{x\nu} \right)
+\left(G_\mu-\frac{g_\mu}{2}-\frac{1}{2}\dot\gamma'\hat{\sigma}_y\delta_{x\mu}\right)
\left(G_\nu-\frac{g_\nu}{2}-\frac{1}{2}\dot\gamma'\hat{\sigma}_y\delta_{x\nu}\right)
 \right\}
\nonumber\\
&\approx&
\Pi^k_{\mu\nu}
\int \frac{d\bm{G}}{\pi^{d/2}}\int \frac{d\bm{g}}{\pi^{d/2}}
\Theta(\bm{g}\cdot\hat{\bm{\sigma}})(\bm{g}\cdot\hat{\bm{\sigma}})^2\hat{\sigma}_\alpha\hat{\sigma}_\beta e^{-2G^2-g^2/2}
\nonumber\\
&&\times
\left[
 2G_\mu G_\nu+\frac{g_\mu g_\nu}{2}
+\frac{1}{2}\dot\gamma' \hat{\sigma}_y\left\{
g_\mu \delta_{x\nu}+g_\nu \delta_{x\mu}-(4G_\mu G_\nu+g_\mu g_\nu)g_x
\right\}
\right]
\nonumber\\
&=&
\frac{2\pi^{d/2}}{d(d+2)\Gamma(d/2)}\Pi^k_{\alpha\beta}.
\end{eqnarray}
Notice that the term proportional to $\dot\gamma'$ in $\tilde{P}_{\alpha\beta}^{c(1)}$ disappears.

The final expression of the collisional pressure tensor can be obtained from Eqs.\ \eqref{P_c(0):result} and \eqref{vic_pc}. It is given by
\begin{equation}\label{P_c:result}
P^c_{\alpha\beta}
=\frac{1+e}{2}n^2\sigma^d g_0 T\frac{\pi^{d/2}}{d\Gamma(d/2)}
\left[
\delta_{\alpha\beta}+\frac{2}{d+2}\Pi^k_{\alpha\beta}
-\dot\gamma' \frac{2\sqrt{2}}{\sqrt{\pi}(d+2)}
\{
\delta_{\alpha x}\delta_{\beta y}+\delta_{\alpha y}\delta_{\beta x}
\}
\right].
\end{equation}

\section{Explicit expressions of $\mathscr{C}_4$, $\mathscr{C}_2$, and $\mathscr{C}_0$}
\label{vicente}

In this Appendix we provide the explicit forms of the quantities $\mathscr{C}_4$, $\mathscr{C}_2$, and $\mathscr{C}_0$. They are given by
\beq
\label{v1}
\mathscr{C}_4 =\mathscr{C}_4^{(0)} \tau_T, \quad \mathscr{C}_4^{(0)}=\frac{d-1}{d}({\cal D}_d+{\cal F}_d)\mathscr{A}_1\tau_T
-\frac{d-2}{d}({\cal E}_d+{\cal F}_d)\mathscr{B}_1\tau_T,
\eeq
\beq
\label{v2}
\mathscr{C}_2=\mathscr{C}_2^{(0)}+\mathscr{C}_2^{(1)}\tau_T,
\eeq
\beq
\label{v3}
\mathscr{C}_0=
\frac{d(2+\nu^*g_0\sqrt{\theta})}{2({\cal C}_d+{\cal F}_d)}\left[g_0 \lambda^*\sqrt{\theta}+2(1-\theta^{-1})\right].
\eeq
Here, we have introduced the auxiliary quantities
\beq
\label{v3.1}
\mathscr{C}_2^{(0)}=\frac{d-1}{d}({\cal D}_d+{\cal F}_d)\mathscr{A}_0- \frac{d-2}{d}({\cal E}_d+
{\cal F}_d)\mathscr{B}_0-{\cal C}_d-\frac{d+2}{2}{\cal F}_d,
\eeq
\beq
\label{v3.2}
\mathscr{C}_2^{(1)}=-\sqrt{\frac{2}{\pi}}
\frac{{\cal F}_d(2+\nu^*g_0\sqrt{\theta})}{{\cal C}_d+{\cal F}_d},
\eeq
\beq
\label{v4}
\mathscr{A}_0=\frac{d(1+ {\cal F}_d)\left[g_0 \gamma^*\sqrt{\theta}+2(1-\theta^{-1})\right]}
{(2+\nu^*g_0 \sqrt{\theta})({\cal C}_d+{\cal F}_d)},
\eeq
\beq
\label{v5}
\mathscr{A}_1=-\sqrt{\frac{2}{\pi}}\frac{2 {\cal F}_d(1-{\cal C}_d)}
{(2+\nu^*g_0 \sqrt{\theta})({\cal C}_d+{\cal F}_d)},
\eeq
\beq
\label{v6}
\mathscr{B}_0=\frac{{\cal E}_d+{\cal F}_d}{1+{\cal F}_d}\mathscr{A}_0, \quad
\mathscr{B}_1=-\frac{1-{\cal C}_d}{{\cal C}_d-{\cal E}_d}\mathscr{A}_1.
\eeq

\section{Outline of the EDLSHS method}
\label{EDLSHS}

In this Appendix, a short outline of the EDLSHS method~\cite{Scala12} under a plane shear \cite{Evans08, Bannerman11} with the aid of the Lees-Edwards boundary condition \cite{LE72} is presented.
The time evolution of $i$-th particle at the position $\bm{r}_i$ and the peculiar momentum  of $i$-th particle are given by Eqs.~\eqref{Langevin_eq} and \eqref{noise}.
The velocity increment from the time $t$ to $t+\Delta t$ in Eqs.~\eqref{Langevin_eq} and \eqref{noise} can be expressed as
\begin{equation}\label{random}
	\bm{v}_{i,\alpha}(t+\Delta t)= e^{-\zeta \Delta t}\bm{v}_{i,\alpha}(t)+\sqrt{\frac{T_{\rm ex}}{m}(1-e^{-2\zeta \Delta t})}
\Xi,
\end{equation}
where $\Xi$ represents a zero mean random number whose variance is 1. In this paper, we use $\Delta t = 0.1/\zeta$ \cite{Scala12}.

To consider the effect of particle collisions, we need to determine the time interval $\Delta \tau$ when the next collision occurs. 
In addition, we also have to detect the events when the particle crosses the Lees-Edwards boundaries at $y=\pm L/2$. 
The time interval $\Delta \tau$ between two sequential events (colliding or crossing the Lees-Edwards boundary) is given by the minimum of the time intervals between (i) the time $\Delta \tau_{ij}>0$ passed for the binary collision of the particles $i$ and $j$, and (ii) the time that the $i$-th particle needs to reach the Lees-Edwards boundary $\Delta \tau_{i,{\rm wall}}>0$. 
While $\Delta \tau_{ij}$ satisfies the condition $|\bm{r}_i (t+\Delta \tau_{ij}) - \bm{r}_j(t+\Delta \tau_{ij})| = \sigma$ in the absence of the random forces, $\Delta \tau_{i,{\rm wall}}$ obeys the condition $y_i (t+\Delta \tau_{i,{\rm wall}})=\pm L/2$ \cite{Bannerman11}. 
Thus, $\Delta \tau$ is determined as $\Delta \tau = \min (\Delta\tau_{ij}, \Delta \tau_{i,{\rm wall}} )$. 
For $t<n\Delta t<t+\Delta \tau$  ($n$ is an integer number), in the absence of collisions, the positions of the particles are updated according to Eq.~\eqref{random}. At $\Delta \tau = \Delta \tau_{ij}$, particles $i$ and $j$ collide and therefore their velocities change according to Eq.~(2), while only the position of the $i$-th particle is updated as $\bm{r}_i \mp \dot\gamma L \Delta t \to \bm{r}_i$ at $\Delta t=\Delta \tau_{i,{\rm wall}}$, where $L$ is the system size and the minus (plus) sign is selected if the velocity is positive (negative).


\section{Results for strong inelastic gas-solid suspensions}
\label{echange}

\begin{figure}[htbp]
\includegraphics[width=170mm]{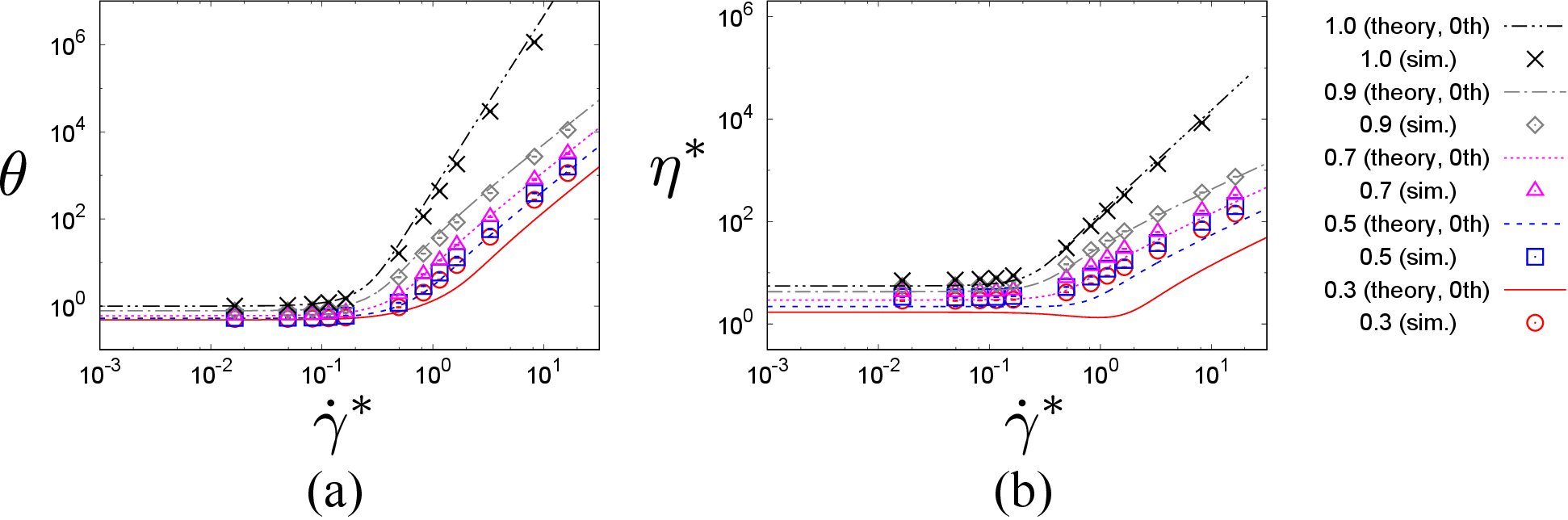}
\caption{
(Color online)
Plots of $\theta$ (panel (a)) and $\eta^*$ (panel (b)) versus the (scaled) shear rate $\dot\gamma^{*}$ for
$\varphi=0.30$ and five different values of the restitution coefficient $e$: $e=1$, $0.9$, $0.7$, $0.5$, and $0.3$.
The lines correspond to the theoretical results obtained from the zeroth-order theory (denoted by 0th in the legend).
Symbols refer to computer simulation results.
  }
\label{e_change}
\end{figure}

\begin{figure}[htbp]
\includegraphics[width=100mm]{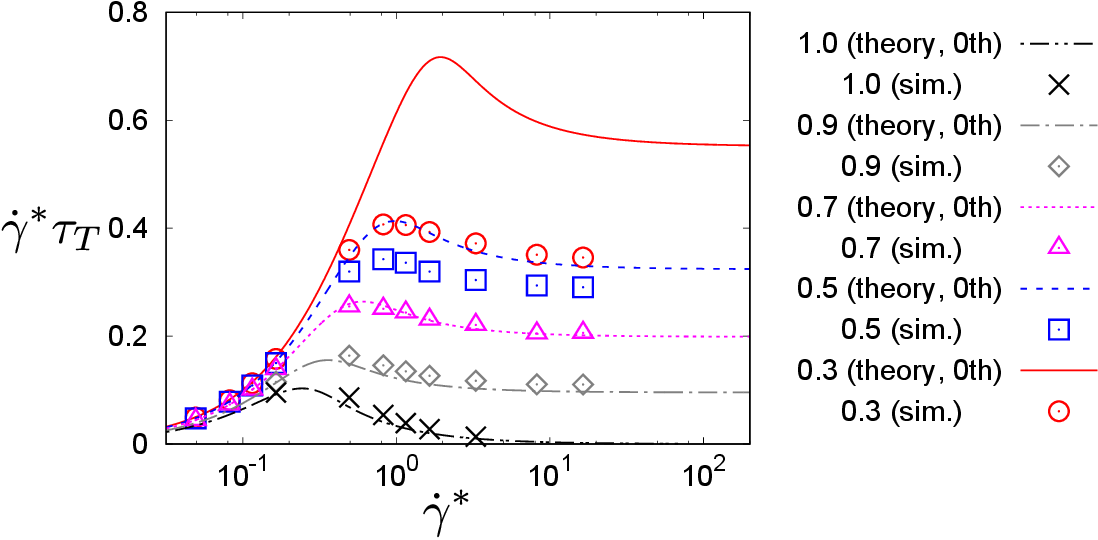}
\caption{
(Color online) Plot of $\dot\gamma^*\tau_T$ versus the (scaled) shear rate $\dot\gamma^{*}$ for $\varphi=0.30$ and five different values of the restitution coefficient $e$: $e=1$, $0.9$, $0.7$, $0.5$, and $0.3$. The lines correspond to the theoretical results obtained from the zeroth-order theory (denoted by 0th in the legend).
Symbols refer to computer simulation results.
}
\label{fig_tauT_e_change}
\end{figure}

As said in Sec.\ \ref{introduction}, in this Appendix we extend our study to suspensions more inelastic than those analyzed in Sec.\ \ref{simulation}. 
More specifically, we present theoretical and simulation results for $\varphi=0.30$ and several values of the restitution coefficient $e$ ($e=1, 0.9, 0.7, 0.5,$ and $0.3$). 
The shear-rate dependence of the kinetic temperature and the shear viscosity of the above systems is plotted in Fig.\ \ref{e_change}. 
Only the results derived from the zeroth-order theory are displayed because they compare better with the simulations than those obtained from the first-order theory as explained in the main text. We also note that the viscosity obtained in the first-order theory becomes negative for $e\le 0.7$.
This unphysical behavior is due to the fact that perturbative parameter $\dot\gamma^*\tau_T$ increases with increasing inelasticity and hence, its contribution in Eq.\ \eqref{P_c:main_text} can be larger enough to lead to a negative value of $\eta^*$. 
It is seen that
the zeroth-order theoretical results for the kinetic temperature $\theta$ and the shear viscosity $\eta^*$ agree well with simulations when  $e\gtrsim 0.5$ and $e\gtrsim 0.7$, respectively. More significant discrepancies are observed for more inelastic systems. Moreover, it is interesting to note that the zeroth-order theory predicts the shear thinning regime near $\dot\gamma^*\simeq0.1$ only for the extreme inelastic case $e=0.3$. 
This feature is not observed in the simulations.



\section{Consistency between kinetic theory and simulations}
\label{app:mu}

\begin{figure}[htbp]
\includegraphics[width=150mm]{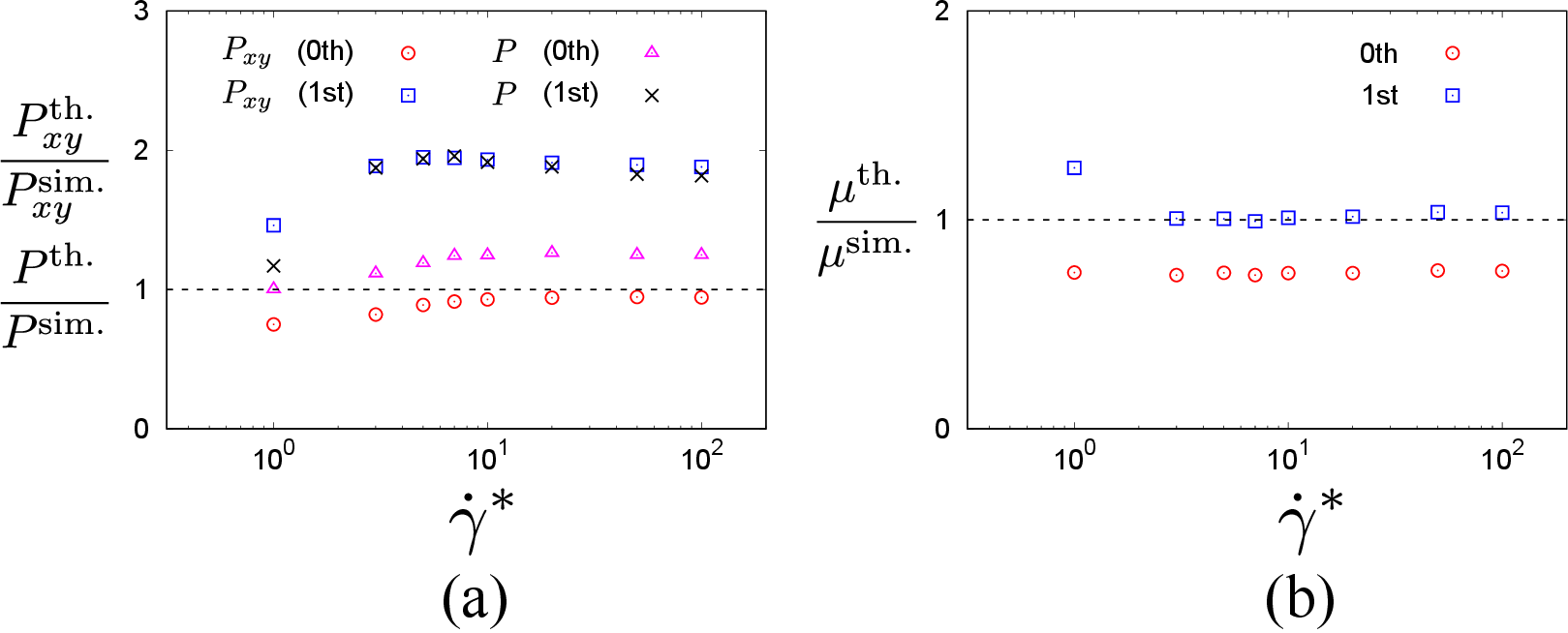}
\caption{
(Color online)
Plots of the ratios of (a) $P_{xy}$ and $P$ and (b) the stress ratio $\mu=-P_{xy}/P$ from the kinetic theory to that from the simulation against $\dot\gamma^*$ for $\varphi=0.30$ with $e=0.9$.
  }
\label{fig_comp}
\end{figure}

In this Appendix, let us check the consistency between the kinetic theory and the simulation for the stress ratio $\mu=-P_{xy}/P$.
Figure \ref{fig_comp} (a) represents the ratios of the predictions of the kinetic theory (see Eqs.~\eqref{P_c:main_text} and \eqref{steady5} as well as $P^k=nT$) for $P_{xy}$ and $P$ to those from the simulation for $\varphi=0.3$.
The approximate results for $\tau_T=0$ from the kinetic theory shows better agreements with those from the simulations than those for $\tau_T>0$ where the approximate results are almost twice larger than the results of simulations.
Nevertheless, the stress ratio $\mu=-P_{xy}/P$ for $\tau_T>0$ becomes better that for $\tau_T=0$ as shown in Fig.~\ref{fig_comp} (b).
This is because both $P_{xy}$ and $P$ for finite $\tau_T$ become twice of the values of the simulation, which gives nearly identical the stress ratio to that of the simulation.

\section{Angular integrals}\label{angular_integral}


First, let us summarize the useful identities which we have already proven:
\begin{equation}\label{Garzo_A5}
\int d\hat{\bm{\sigma}}\Theta(\bm{c}\cdot\hat{\bm{\sigma}})(\bm{c}\cdot\hat{\bm{\sigma}})^n=B_n c^n ,
\end{equation}
\begin{equation}\label{Garzo_A6}
\int d\hat{\bm{\sigma}}\Theta(\bm{c}\cdot\hat{\bm{\sigma}})(\bm{c}\cdot\hat{\bm{\sigma}})^n \hat{\sigma}_\alpha
=B_{n+1}c^{n-1}c_\alpha ,
\end{equation}
\begin{equation}\label{Garzo_A17}
\int d\hat{\bm{\sigma}}\Theta(\bm{c}\cdot\hat{\bm{\sigma}})(\bm{c}\cdot\hat{\bm{\sigma}})^n \hat{\sigma}_\alpha\hat{\sigma}_\beta
=\frac{B_n}{n+d}c^{n-2}(n c_\alpha c_\beta+c^2 \delta_{\alpha\beta})
\end{equation}
where
\begin{equation}\label{beta}
B_n=\pi^{(d-1)/2}\frac{\Gamma(\frac{n+1}{2})}{\Gamma(\frac{n+d}{2})} ,
\quad
B_n^{d=3}=\frac{2\pi}{n+1} .
\end{equation}
We often use the area of the hyper-unit sphere in $d-$dimension
\begin{equation}\label{S_d}
S_d=\frac{d\pi^{d/2}}{\Gamma(d/2+1)}=\frac{2\pi^{d/2}}{\Gamma(d/2)} .
\end{equation}

Using these identities, we can prove
\begin{equation}\label{PoF_B14}
\int d\hat{\bm{\sigma}}\Theta(\hat{\bm{\sigma}}\cdot\bm{c})(\hat{\bm{\sigma}}\cdot\bm{c})^n\hat{\sigma}_\alpha\hat{\sigma}_\beta \hat{\sigma}_\gamma
=\frac{B_{n+1}}{n+d+1} c^{n-3} [(n-1)c_\alpha c_\beta c_\gamma+
c^2( \delta_{\alpha\beta}c_\gamma +
\delta_{\alpha\gamma}c_\beta+\delta_{\beta\gamma}c_\alpha) ]
\end{equation}
and
\begin{eqnarray}\label{PoF_B15}
\int d\hat{\bm{\sigma}}\Theta(\hat{\bm{\sigma}}\cdot\bm{c})(\hat{\bm{\sigma}}\cdot\bm{c})^2\hat{\sigma}_\alpha\hat{\sigma}_\beta \hat{\sigma}_\gamma\hat{\sigma}_\delta
&=&
\frac{B_2}{(d+2)(d+4)}
[2(c_\alpha c_\beta \delta_{\gamma\delta}+c_\alpha c_\gamma \delta_{\beta\delta}+c_{\alpha} c_{\delta}\delta_{\beta\gamma}+c_{\beta}c_{\gamma}\delta_{\alpha\delta}+
c_{\beta}c_{\delta}\delta_{\alpha\gamma}+c_{\gamma}c_{\delta}\delta_{\alpha\beta})
\nonumber\\
&&
+c^2(\delta_{\alpha\beta}\delta_{\gamma\delta}+\delta_{\alpha\gamma}\delta_{\beta\delta}
+\delta_{\alpha\delta}\delta_{\beta\gamma})]
\end{eqnarray}
for a positive integer $n$~\cite{Suppl}.


\end{document}